\documentclass[aps, 10pt, prd, twocolumn, superscriptaddress, preprintnumbers, longbibliography, nofootinbib]{revtex4-1}

\usepackage[top=2.8cm, bottom=2.8cm, left=2cm, right=2cm]{geometry}
\usepackage[utf8]{inputenc} 
\usepackage[T1]{fontenc}
\usepackage{amsmath, amssymb, lmodern} 
\usepackage[dvipsnames]{xcolor}
\usepackage{graphicx}
\usepackage{subfigure}
\usepackage{natbib}
\usepackage{booktabs}
\usepackage{placeins}
\usepackage{color} 
\usepackage{hyperref}
\usepackage{tabularray}
\hypersetup{colorlinks, citecolor = blue, filecolor = blue, linkcolor = blue, urlcolor = blue}
\usepackage[normalem]{ulem}
\usepackage{comment}

\definecolor{darkgreen}{rgb}{0, 0.6, 0}

\newcommand{\dd}{{\text{d}}}
\newcommand{\Mpc}{\text{Mpc}}
\newcommand{\MG}{\text{MG}}
\newcommand{\class}{{\texttt{CLASS}}}
\newcommand{\nw}{{\text{nw}}}
\newcommand{\w}{{\text{w}}}
\newcommand{\eff}{\text{eff}}
\newcommand{\eq}{\text{eq}}
\renewcommand{\max}{\text{max}}
\newcommand{\Silk}{\text{Silk}}
\renewcommand{\S}{\text{S}}
\newcommand{\Acc}{\text{Acc}}
\newcommand{\mape}{\text{MAPE}}
\newcommand{\GA}{\text{GA}}
\newcommand{\GAnw}{{\text{GA},\text{nw}}}
\newcommand{\EH}{\text{EH}}
\newcommand{\SG}{\text{SG}}

\begin{document}

\preprint{IFT-UAM/CSIC-24-110}

\title{An Interpretable and Physics-Informed Emulator for the Linear Matter Power Spectrum from Machine Learning}

\author{J. Bayron Orjuela-Quintana}
\email{john.orjuela@correounivalle.edu.co}
\affiliation{Departamento  de  F\'isica,  Universidad  del Valle, Ciudad  Universitaria Mel\'endez,  Santiago de Cali  760032,  Colombia}
\affiliation{Cosmology and Theoretical Astrophysics group, Departamento de F\'isica, FCFM, Universidad de Chile, Blanco Encalada 2008, Santiago, Chile}

\author{Domenico Sapone}
\email{domenico.sapone@uchile.cl}
\affiliation{Cosmology and Theoretical Astrophysics group, Departamento de F\'isica, FCFM, Universidad de Chile, Blanco Encalada 2008, Santiago, Chile}

\author{Savvas Nesseris}
\email{savvas.nesseris@csic.es}
\affiliation{Instituto  de  F\'isica  Te\'orica  UAM-CSIC,  Universidad  Auton\'oma  de  Madrid,  Cantoblanco,  28049  Madrid,  Spain}

\begin{abstract}

We present an interpretable emulator for the linear matter power spectrum (MPS) in the standard cosmological model $\Lambda$CDM, constructed via a \textit{physics-informed symbolic regression} framework. By combining domain knowledge with a machine learning technique known as genetic algorithms, we explore the space of analytic expressions to derive closed-form, smooth, physically motivated approximations of the MPS that match the accuracy of standard broadband reconstruction methodologies such as the Savitzky-Golay filter. Building upon this baseline, we incorporate transparent oscillatory corrections informed by the physics of baryon acoustic oscillations (BAO). The resulting expression delivers mean sub-percent fractional errors across a broad range of scales ($k \in [10^{-5}, 1.5]~h\,\mathrm{Mpc}^{-1}$) with an average deviation of $\sim 0.4\%$ when tested against spectra computed with a Boltzmann solver. Moreover, a comparable level of fractional deviation is maintained on smaller scales when the GA-derived formulation is used as input to the nonlinear emulator \texttt{halofit}. To illustrate the versatility of the framework beyond $\Lambda$CDM, we apply it to a representative $f(R)$ gravity model. Rather than training a general modified-gravity emulator, we compute the corresponding linear spectra with a Boltzmann solver and fit a parametric deformation of the $\Lambda$CDM smoothed component. This procedure achieves average errors at the 1.5–1.8\% level and captures the leading modulation of the MPS induced by modified gravity, enabling a controlled study of its impact on the BAO scale. Our results provide compact, accurate, and physically motivated fitting functions for the linear MPS in both standard and MG cosmologies, offering a fast and transparent alternative to existing emulators for parameter inference and theoretical modeling in large-scale structure analyses.

\end{abstract}

\maketitle

\section{Introduction} 
\label{Sec: Intro}

The matter power spectrum (MPS), $P(k)$, plays a central role in modern cosmology. It encapsulates the statistical distribution of matter density fluctuations in the early universe and serves as the foundational ingredient for predicting a wide range of observables in the large-scale structure (LSS) of the Universe~\cite{Dodelson:2020bqr}. As linear perturbations grow and evolve under gravity, $P(k)$ informs key cosmological signatures, including galaxy clustering~\cite{SDSS:2003tbn}, weak gravitational lensing~\cite{Kilbinger:2014cea}, and redshift-space distortions~\cite{Kaiser:1987qv}. In particular, the position and shape of the baryon acoustic oscillation (BAO) features imprinted in the MPS provide a robust cosmic standard ruler, enabling precise measurements of the expansion history~\cite{BOSS:2016wmc}. Current and upcoming surveys such as DESI~\cite{DESI:2016fyo}, Euclid~\cite{EUCLID:2011zbd}, LSST~\cite{LSSTScience:2009jmu}, and the Nancy Grace Roman Space Telescope~\cite{Spergel:2015sza} rely critically on accurate predictions of $P(k)$ across a wide range of scales and cosmological parameters to constrain the physics of the dark sector and test deviations from the standard $\Lambda$CDM model.

Boltzmann solvers like \class~\cite{Blas:2011rf} and \texttt{CAMB}~\cite{Lewis:1999bs} provide high-precision computations of the linear MPS, but could become computationally expensive when deployed in large-scale inference pipelines. These pipelines typically require thousands of evaluations across high-dimensional parameter spaces, as in Markov Chain Monte Carlo (MCMC) methods. To alleviate this computational cost, fast emulators have been developed using machine learning techniques, including neural networks and Gaussian process regressions trained on dense grids of precomputed spectra. Representative examples include the \texttt{Euclid Emulator}~\cite{Euclid:2018mlb}, \texttt{CosmoPower}~\cite{SpurioMancini:2021ppk}, and \texttt{BACCO}~\cite{Arico:2020lhq}. While such methods offer substantial speed gains and flexibility, they often sacrifice physical transparency and require large training sets to achieve competitive accuracy. Moreover, modifying these models to incorporate new physics---such as effects from modified gravity (MG) theories~\cite{Clifton:2011jh}, massive neutrinos~\cite{Lesgourgues:2006nd}, or dynamical dark energy~\cite{Copeland:2006wr}---usually entails costly retraining and risks obscuring the physical interpretation of results.

These limitations motivate the exploration of alternative approaches that balance speed, accuracy, and interpretability. Symbolic regression (SR) offers a promising strategy in this direction, aiming to discover explicit mathematical expressions that fit data while retaining physical transparency~\cite{Cranmer:2020wew}. Recent advances have led to the development of powerful tools such as \texttt{AI Feynman}~\cite{Udrescu:2019mnk}, which combines neural networks and physics-inspired priors to rediscover known laws of nature, and \texttt{PySR}~\cite{Cranmer:2023ar}, a high-performance symbolic regression engine using evolutionary algorithms to explore the mathematical expressions space. Other SR tools include \texttt{gplearn} (a scikit-learn-compatible GP library~\cite{gplearn}), \texttt{TuringBot} (a commercial SR engine~\cite{turingbot}), and neural-symbolic approaches such as \texttt{NSR}~\cite{Petersen:2021deep}.

Cosmology, in parallel, has a long-standing tradition of developing fitting formulas for $P(k)$. One of the earliest and most influential examples is the BBKS formula~\cite{Bardeen:1985tr}, which provided one of the first widely adopted semi-analytic approximations for cold dark matter (CDM) transfer functions. About a decade later, Eisenstein and Hu introduced their well-known fitting functions~\cite{Eisenstein:1997ik, Eisenstein:1997jh}, delivering accurate and physically motivated templates for the baryon acoustic oscillation (BAO) features in the matter transfer function—tools that remain in extensive use today. More recently, symbolic emulators such as \texttt{symbolic\_pofk}~\cite{Bartlett:2023cyr, Bartlett:2024jes, Sui:2024wob} have demonstrated the power of SR to model the linear and nonlinear MPS across a wide parameter range, including scenarios beyond $\Lambda$CDM.

In this work, we present an interpretable emulator for the linear MPS, constructed using a physics-informed symbolic regression strategy. Our approach leverages a machine learning technique knows as genetic algorithms (GAs)—evolutionary optimization methods inspired by natural selection~\cite{Koza:1992gen}—to search efficiently over a space of analytic expressions. To guide this search, we incorporate domain-specific physical priors related to the known scaling of the transfer function and the oscillatory structure of BAO. This prior knowledge restricts the model space to physically meaningful candidates, enhancing interpretability and reducing overfitting. This approach provides analytic transparency while retaining enough flexibility to accommodate localized empirical corrections required for better performance.

We draw an analogy between MCMC sampling in cosmological parameter spaces and the stochastic exploration performed by GAs in model space. Just as MCMC uses priors to constrain the region of exploration~\cite{Trotta:2008qt, Trotta:2017wnx}, we apply physical intuition to confine the symbolic search. This strategy distinguishes our method from more general-purpose tools like \texttt{AI Feynman}~\cite{Udrescu:2019mnk, Udrescu:2020ai}, which seek symmetries across arbitrary datasets. Instead, our method targets a specific physical phenomenon, enabling the construction of transparent, compact formulas that expose how cosmological parameters influence the shape of the MPS.

Building upon this physically motivated foundation, we propose a parametric correction scheme designed to model typical deviations from $\Lambda$CDM as predicted by MG scenarios. These corrections are expressed as multiplicative deformations to the non-wiggly component of $P(k)$, enabling us to capture key signatures such as the suppression or enhancement of power at specific scales. While not intended as a full emulator for arbitrary MG models, our framework provides a versatile tool to isolate and study the impact of representative MG-induced effects on observable quantities.

This paper is organized as follows. Section~\ref{Sec: GAs} provides a concise review of GAs and SR. In Section~\ref{Sec: Matter Power Spectrum}, we introduce the theoretical background underlying the linear MPS. Section~\ref{Sec: Emulator for LCDM} outlines the data generation pipeline and presents the physics-informed SR strategy used to derive our fitting formulas. Then, we compare the complexity and accuracy of our resulting symbolic formulation against existing approaches, highlighting both the strengths and limitations of our methodology. In Section~\ref{Sec: Parametric Formula for MG Theories}, we introduce parametric extensions to describe typical MG effects. Section~\ref{Sec: BAO Scale} uses the resulting models to quantify shifts in the BAO scale under the influence of a representative MG model. We conclude in Section~\ref{Sec: Conclusions} with a summary and outlook.

\section{Symbolic Regression and Genetic Algorithms} 
\label{Sec: GAs}

Genetic programming (GP) is a machine learning paradigm inspired by biological evolution. It evolves symbolic structures—such as mathematical expressions—through natural selection, crossover, and mutation~\cite{Holland:1984gen}, optimizing them according to a fitness criterion that typically quantifies how well a candidate model reproduces the target data. This evolutionary approach is particularly suited for SR, where the goal is to uncover interpretable analytical relationships between inputs and outputs.

In GP-based SR, the algorithm begins with a randomly generated population of symbolic expressions, often represented as expression trees built from a user-defined set of functions (e.g., polynomials, trigonometric functions, exponential functions) and operations (e.g., addition, multiplication, composition). Each expression is evaluated using a fitness function—commonly the mean squared error. The fittest individuals are selected to produce the next generation via genetic operations:
\begin{itemize}
\item \textbf{Crossover:} Subtrees from two parent expressions are swapped to recombine components.
\item \textbf{Mutation:} A random subexpression is replaced or modified, introducing variation.
\item \textbf{Elitism:} Top-performing individuals are preserved to retain progress.
\end{itemize}
This evolutionary cycle continues until convergence or until a predefined computational budget is exhausted. The final output is a compact, human-readable formula that, in principle, balances accuracy with simplicity, often revealing physically meaningful structures.

GAs have found widespread applications in the physical sciences, from approximating solutions to the Schrödinger equation~\cite{Lahoz-Beltra:2022ser} and classifying Calabi-Yau manifolds~\cite{Berglund:2023ztk} to gravitational wave detection~\cite{Gupte:2020mfp} and astrophysics and cosmology~\cite{Luo:2019qbk, Arjona:2019fwb, Arjona:2020kco, Alestas:2022gcg, Medel-Esquivel:2023nov, Bogdanos:2009ib, Nesseris:2010ep, Arjona:2021hmg, Arjona:2021mzf, Arjona:2020axn, Euclid:2020ojp, Euclid:2021frk}. Their flexibility in high-dimensional, nonlinear settings makes them particularly attractive for modeling complex physical systems. 

In this work, we use a customized GA-based SR code\footnote{\label{FN: GA code}All codes developed in this work are publicly available at:\\
\href{https://github.com/BayronO/Pk-Emulator-from-GAs}{\texttt{https://github.com/BayronO/Pk-Emulator-from-GAs}}.\\
The GA implementation builds upon the framework provided by one of the authors, available at:\\
\href{https://github.com/snesseris/Genetic-Algorithms}{\texttt{https://github.com/snesseris/Genetic-Algorithms}}.} tailored to our cosmological application. We also employ \texttt{PySR}~\cite{Cranmer:2020wew}, a modern SR library that leverages just-in-time (JIT) compilation, parallelization, and multi-objective optimization. Crucially, we incorporate domain knowledge—e.g., restricting the functional space to expressions consistent with physical scalings or symmetries—into our code. This ``\textit{physics-informed}'' strategy restricts the search space, promotes interpretability, and avoids overfitting, yielding compact expressions that remain accurate across the cosmological parameter space.

\section{The Linear Matter Power Spectrum} 
\label{Sec: Matter Power Spectrum}

On cosmological scales exceeding approximately 100 Mpc, the Universe appears remarkably homogeneous and isotropic, as confirmed by observations of the cosmic microwave background and large-scale galaxy surveys~\cite{Yadav:2010}. On smaller scales, however, it exhibits significant inhomogeneities—the cosmic web of galaxies, clusters, and filaments—formed through the gravitational collapse of matter~\cite{Bond:1995yt}. The statistical distribution of these structures is encoded in the MPS, $P(k)$, which quantifies the variance of matter density fluctuations as a function of the comoving wavenumber $k$, typically expressed in units of $h\,\Mpc^{-1}$, where $h$ is the dimensionless Hubble parameter~\cite{Dodelson:2020bqr}.

In the standard cosmological $\Lambda$CDM model, the late-time gravitational potential arises from primordial curvature perturbations generated during inflation. This connection is mediated by two key functions:
\begin{enumerate}
\item[$i)$] the matter transfer function (MTS) $T(k)$, describing the scale-dependent evolution of perturbations from super-horizon scales through horizon crossing and the transition from radiation to matter domination;
\item[$ii)$] the linear growth factor $D_+(a)$, characterizing the time evolution of matter overdensities in the linear regime, where $a$ is the scale factor.
\end{enumerate}

Assuming adiabatic initial conditions and linear theory, the gravitational potential in Fourier space evolves as:
\begin{equation}
\Phi(k, a) \propto T(k) D_+(a) \sqrt{\Delta_\mathcal{R}(k)},
\end{equation}
where $\Delta_\mathcal{R}(k)$ is the dimensionless primordial curvature power spectrum, usually parametrized as:
\begin{equation}
\Delta_\mathcal{R}(k) = A_s \left(\frac{k}{k_{\mathrm{p}}}\right)^{n_s - 1},
\end{equation}
with $A_s$ the amplitude of scalar perturbations, $n_s$ the spectral index, and $k_\text{p} = 0.05~\text{Mpc}^{-1}$ the pivot scale adopted by Planck~\cite{Planck:2018vyg}.

At late times and for sub-horizon modes, the gravitational potential relates to the matter overdensity field $\delta_m$ via the Poisson equation:
\begin{equation}
k^2 \Phi(k, a) \propto a^2 \bar{\rho}_m (a) \delta_m(k, a),
\end{equation}
where $\bar{\rho}_m(a)$ is the background density of pressureless matter. Assuming nearly Gaussian, zero-mean primordial perturbations, the linear MPS becomes:
\begin{equation}
\label{Eq: General MPS}
    P(k, a) \propto \frac{2\pi^2}{k^3} \left[\frac{k^4}{\bar{\rho}_m^2(a)} P_\mathcal{R}(k) D_+^2(a) T^2(k)\right].
\end{equation}
Evaluated at the present epoch ($a = 1$), this simplifies to:
\begin{equation}
P(k) \propto k^{n_s} T^2(k),
\end{equation}
highlighting that the present-day MPS is fully determined by the shape of the MTS and the primordial spectral tilt. This expression encapsulates how initial density fluctuations evolve into the observed LSS, with $T(k)$ encoding the relevant physical processes.

The physics shaping the MPS is rich and multifaceted. While CDM drives gravitational clustering, baryons introduce additional features. Prior to recombination, baryons were tightly coupled to photons, resulting in acoustic oscillations due to radiation pressure. These BAOs imprint a characteristic modulation on $P(k)$ at intermediate scales, analogous to features in the CMB angular power spectrum~\cite{Eisenstein:1997ik}.

Beyond the standard CDM+baryon picture, other physical effects can significantly alter $P(k)$. For example, massive neutrinos suppress small-scale structure growth due to their thermal velocities and free-streaming behavior~\cite{Agarwal:2011}. Similarly, dark energy or MG theories affect the evolution of matter perturbations by: $(i)$ altering the expansion rate and thus the growth of fluctuations, and $(ii)$ introducing additional clustering or modifying the gravitational interaction itself~\cite{Lombriser:2015axa}.

\subsection{Calculation of Matter Power Spectra}
\label{Sec: Calculation of MPS}

The process of structure formation involves a complex interplay of gravitational dynamics and various microphysical processes, making it exceedingly challenging to derive complete analytical solutions from the set of nonlinear Einstein-Boltzmann equations. Nevertheless, efficient numerical solutions can be achieved using advanced Boltzmann solvers, as demonstrated by software packages such as  \class~\cite{Blas:2011rf} and \texttt{CAMB}~\cite{Lewis:1999bs}. These codes, which by default assume the concordance model, are highly modular and allow for the incorporation of extended models. Notable examples include MG theories, non-standard dark energy dynamics~\cite{Zhao:2008bn, Zucca:2019xhg, Wang:2023tjj, Zumalacarregui:2016pph, Sakr:2021ylx}, warm dark matter scenarios, and interacting dark sector components~\cite{Becker:2020hzj}.

Beyond linear order, the most accurate method for extracting theoretical insights about MPS stems from large $N$-body simulations, such as the \texttt{Quijote} suite~\cite{Villaescusa-Navarro:2019bje}. Nonetheless, a significant drawback of this approach is its substantial computational burden and limited adaptability to incorporate new models. Recent advancements have addressed the computational challenges by employing emulation techniques like neural networks or Gaussian processes to analyze $P(k)$ data from these simulations~\cite{Heitmann:2013bra, Angulo:2020vky, Mootoovaloo:2021rot, SpurioMancini:2021ppk}. Although emulated MPS offer high accuracy and rapid computation, it is worth noting that they suffer from the same limited adaptability which inherits from $N$-body simulations.

An alternative approach lies in semi-analytical formulations, which provide several practical and conceptual advantages over purely data-driven emulators. Although deep learning models often achieve superior predictive accuracy, semi-analytical methods are more transparent, easier to integrate into existing computational pipelines, and free from external dependencies such as emulator libraries or installation overheads. More importantly, as will be discussed in subsequent sections, these formulations can incorporate physically motivated priors that guide the model toward functionally interpretable solutions. This capability enhances scientific insight and contrasts with the opaque, black-box nature of many deep learning models.

Owing to the distinct physical origin of the acoustic oscillations in the MPS, it is possible to decompose the full signal into a smooth, broad-band component and a superimposed oscillatory feature associated with BAOs. This separation facilitates the modeling of small oscillatory fluctuations on top of a dominant background, thereby simplifying analytical exploration. In the following section, we pursue a semi-analytical representation of the linear MPS considering this separation.

\section{Emulator for the $P(k)$ of $\Lambda$CDM} 
\label{Sec: Emulator for LCDM}

As motivated in the previous section, the linear MPS can be effectively modeled by separating the MTS into two components:
\begin{equation}
    T(k) \equiv T_\nw(k) \, T_\w(k),
\end{equation}
where $T_\nw(k)$ encodes the smooth, broadband evolution of perturbations, while $T_\w(k)$ captures the oscillatory modulation arising from BAOs. The smooth component $T_\nw(k)$ is typically extracted through a ``de-wiggling'' procedure applied to the full MTS. Common strategies include evaluating the Eisenstein--Hu (EH) fitting formula in the zero-baryon limit~\cite{Eisenstein:1997ik, eBOSS:2020abk}, or applying digital filtering techniques—such as the Savitzky–Golay (SG) filter~\cite{Hinton:2016atz, Euclid:2023tqw}—to remove the oscillatory structure.

In what follows, we develop a semi-analytical emulator for $P(k)$ by constructing compact symbolic expressions for both $T_\nw(k)$ and $T_\w(k)$ using GAs. These expressions are informed by physical priors and calibrated on numerical data from Boltzmann solvers, combining interpretability with precision across the cosmological parameter space.

\subsection{The De-Wiggled Matter Power Spectrum}
\label{Sec: The De-Wiggled Matter Power Spectrum}

While deriving a complete analytical description of $P(k)$ from first principles remains unattainable, it appears feasible to capture the typical ``mountain'' shape of the MPS using a simpler formulation. In the following, we focus on constructing a compact and accurate representation of this characteristic broad-band structure, which forms the backbone of the full MPS.

\subsubsection{\textbf{Training Data}}

To construct the dataset for modeling the smooth MTS $T_\nw(k)$, we sample a $4 \times 4 \times 4$ grid over the cosmological parameters $\{h, \omega_b, \omega_m\}$, where $\omega_b$ and $\omega_m$ denote the reduced density parameters of baryons and total matter, respectively. For each point in this parameter grid, we use \class~to compute the linear gravitational potential $\Phi(k)$ over 114 logarithmically spaced $k$-values, spanning the range $k \in [10^{-5}, 1.5]~h\,\Mpc^{-1}$. These are the scale range and number of points in which \class~computes $\Phi(k)$ by default. An overview of the sampled parameter ranges is provided in Table~\ref{Tab: Params}.

The MTS is then obtained by normalizing the potential with respect to its value at the largest scale:
\begin{equation}
T(k) \equiv \frac{\Phi(k)}{\Phi(k_\text{min})},
\end{equation}
where $k_\text{min}$ is the smallest $k$-value used by \class. This procedure yields a dataset of shape $\{64 \times 114, 5\}$, organized as tuples of the form $\{k, h, \omega_b, \omega_m, T(k)\}$. 

To facilitate the search for a compact symbolic model, we remove 44 high-$k$ data points from each of the 64 parameter sets, restricting attention to the domain $k > 0.05~h\,\Mpc^{-1}$. This cutoff reduces contamination from BAOs, which are not part of the smooth spectrum and could hinder the generalization of SR. The resulting training set consists of $4480$ entries—remarkably compact compared to conventional emulators, which often require tens or hundreds of thousands of data points.

\begin{center}
\begin{table}
\begin{centering}
\begin{tblr}{c | c | c}
\hline[1.2pt]
\hline
Variable & Min Value & Max Value \\
\hline
\hline[1.2pt]
$h$ & 0.65 & 0.75 \\
\hline
$\omega_b$ & 0.0214 & 0.0234 \\
\hline
$\omega_m$ & 0.13 & 0.15 \\
\hline
$n_s$ & 0.9 & 1.0 \\
\hline
$A_s \times 10^9$ & 1.5 & 2.5 \\
\hline
$k / \left(h\,\Mpc^{-1}\right)$ & $10^{-5}$ & 1.5 \\
\hline[1.2pt]
\end{tblr}
\par\end{centering}
\caption{Cosmological parameter ranges used to generate the training and test datasets. These values span a region extending several standard deviations around the Planck 2018 best-fit $\Lambda$CDM parameters~\cite{Planck:2018vyg}. The range for $k$ corresponds to the by default range considered by \class.}
\label{Tab: Params}
\end{table}
\par\end{center}

\subsubsection{\textbf{Template for the GA}}

Before the development of modern Boltzmann solvers, several analytical approximations were proposed to model the MTS. Two of the most influential are the zero-baryon limit of the EH model and the BBKS formula. Both aim to reproduce the shape of $T(k)$ in the limit where baryonic effects can be neglected.

The zero-baryon EH formula is a simplified version of the full fitting function presented in Ref.~\cite{Eisenstein:1997ik}. It captures the suppression of power at small scales due to horizon crossing during radiation domination. It reads:
\begin{equation}
\label{Eq: TEHnw}
    T_{\EH, \nw} (k) = \frac{L_0}{L_0 + C_0 q_\EH^2},
\end{equation}
where:
\begin{align}
    L_0 &\equiv \log(2e + 1.8 q_\EH), \\
    C_0 &\equiv 14.2 + \frac{731}{1 + 62.5 q_\EH}, \\
    q_\EH &\equiv \frac{k}{h}\frac{\Theta^2}{\Gamma_\eff}, \\
    \Gamma_\eff &\equiv \frac{\omega_m}{h}\left( \alpha_\Gamma + \frac{1 - \alpha_\Gamma}{1 + (0.43 k s_\EH)^4}\right), \\
    \alpha_\Gamma &\equiv 1 - 0.328 \log(431.0\omega_m)\left( \frac{\omega_b}{\omega_m} \right) \nonumber \\
    &+ 0.38 \log(22.3 \omega_m) \left( \frac{\omega_b}{\omega_m} \right)^2, \\
    s_\EH &\equiv \frac{44.5 \log(9.83/\omega_m)}{\sqrt{1 + 10\omega_b^{3/4}}}~[\Mpc], \label{Eq: sEH}
\end{align}
with $T_\text{CMB} = 2.7\Theta$ K the present-day CMB temperature. This expression improves upon earlier approximations by satisfying the expected $k^2$ suppression at large scales dictated by causality~\cite{Zeldovich:1965gev}. 

A simpler yet historically important alternative is the BBKS fitting function~\cite{Bardeen:1985tr}, valid in the limit $\Omega_b \ll \Omega_m$. It is given by:
\begin{align}
    &T_\text{BBKS}(k) \equiv \frac{\ln(1 + 2.34 q_\text{B})}{2.34 q_\text{B}} \Big[ 1 + 3.89 q_\text{B} \nonumber \\
    &+ (16.1 q_\text{B})^2 + (5.46 q_\text{B})^3 + (6.71 q_\text{B})^4 \Big]^{-1/4},
\end{align}
with
\begin{equation}
q_\text{B} \equiv \frac{k~\theta^{1/2}}{\omega_m - \omega_b}, \qquad \theta \equiv \frac{\rho_r}{1.68 \rho_\gamma},
\end{equation}
where $\rho_r$ and $\rho_\gamma$ denotes the radiation and photon density, respectively.

Although the BBKS formula lacks the precision required for modern cosmological analyses, it successfully captures the essential “mountain-like” shape of the MPS. It is also compact, smooth, and satisfies important asymptotic properties:
\begin{itemize}
    \item[$\bullet$] $\displaystyle\lim_{k \rightarrow 0} T_\text{BBKS} = 1$, \quad $\displaystyle\lim_{k \rightarrow \infty} T_\text{BBKS} = 0$;
    \item[$\bullet$] Non-negativity: $T_\text{BBKS}(k) \geq 0$ for all $k$;
    \item[$\bullet$] Smooth and easy to evaluate.
\end{itemize}

Inspired by this structure, we define our SR template as:
\begin{equation}
T_\GAnw(q) \equiv \left[ 1 + \sum_{i=1}^{6} a_iq^{b_i} \right]^{-1/4},
\end{equation}
where
\begin{equation}
q(k|h, \omega_b, \omega_m) \equiv \frac{hk}{\omega_m - \omega_b},
\end{equation}
and the parameters $\{a_i, b_i\}$ are to be optimized by the GA. This fixed-form ansatz balances expressiveness and complexity, preventing overfitting while retaining flexibility to model $T(k)$ across the sampled cosmological space.

The fitness of any candidate expression $T_\text{expr}(k)$ is evaluated through the mean absolute percentage error (\mape):
\begin{equation}
\label{Eq: Acc GA}
    \mape \equiv \frac{1}{N}\sum_{i = 1}^N \left| \frac{T_{i, \class} - T_{i, \text{expr}}}{T_{i, \class}} \right| \times 100,
\end{equation}
where $N$ is the number of sampled points. For more implementation details of the GA code, please refer to Footnote~\ref{FN: GA code}.

Finally, we quantify the complexity of an analytical expression using two metrics: the \textit{leaf count} [$L(\text{expr})$], which measures the number of basic components (i.e., constants, variables, and operations) required to represent the expression, and the \textit{depth} [$D(\text{expr})$], which corresponds to the number of layers in its syntactic tree. For example, the expression $x^3$ has a leaf count $L(x^3) = 3$, as it consists of the variable $x$, the constant $3$, and the exponentiation operation. Its depth is $D(x^3) = 2$, since the tree structure has two levels: the exponentiation node connects the base and the exponent, forming the structure $x \leftarrow \texttt{pow} \rightarrow 3$.

\subsubsection{\textbf{Fitting Formula}}

\begin{figure*}[t!]
\centering    
\includegraphics[width=\columnwidth]{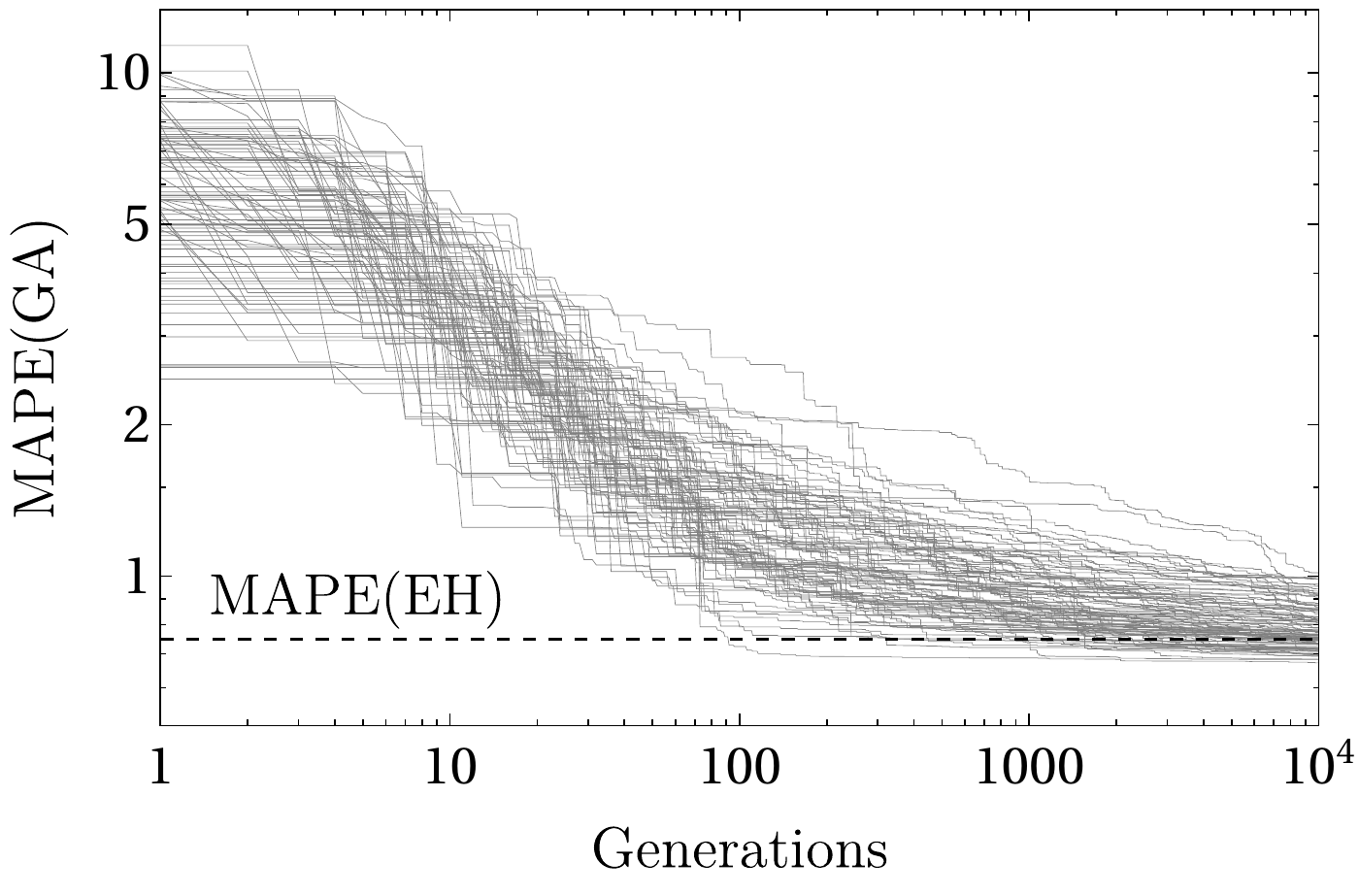} \hfill
\includegraphics[width=\columnwidth]{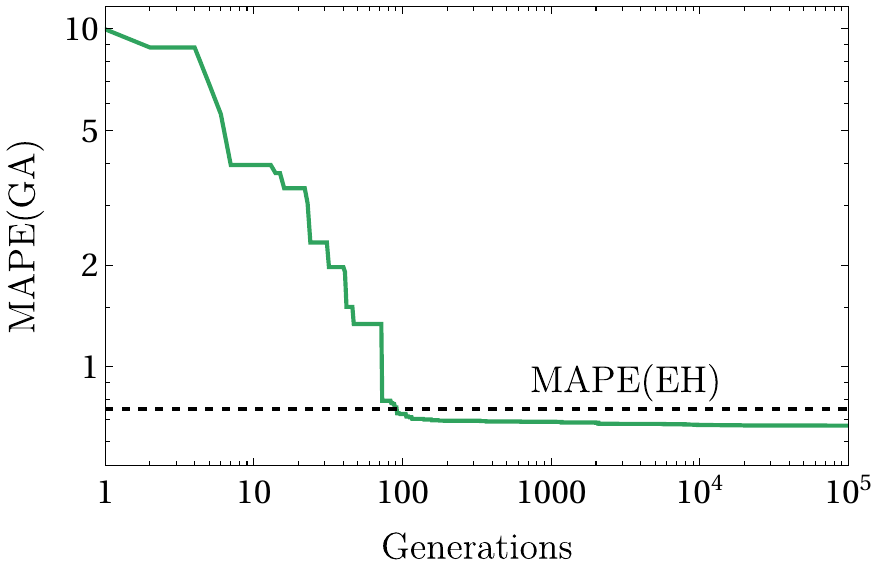}
\caption{\textbf{Left}: Evolution of the fitness function across $10^4$ generations for 100 different GA runs initialized with different random seeds. \textbf{Right}: Fitness evolution for the best-performing run extended to $10^5$ generations. The plateau indicates stagnation in the optimization process.}
\label{Fig: Seeds}
\end{figure*}

\begin{center}
\begin{table}
\begin{centering}
\begin{tblr}{ c  c | c  c }
\hline[1.2pt]
\hline
Constant & Value & Constant & Value \\
\hline
\hline[1.2pt]
$a_1$ & 101.855 & $b_1$ & 1.483 \\
\hline
$a_2$ & 21112.189 & $b_2$ & 3.972 \\
\hline
$a_3$ & 35913.065 & $b_3$ & 6.097 \\
\hline
$a_4$ & 1428.081 & $b_4$ & 7.507 \\
\hline[1.2pt]
\end{tblr}
\caption{Coefficients of the final GA-derived fitting formula for the no-wiggle transfer function $T_{\GA, \nw(k)}$.}
\label{Tab: Tnw}
\end{centering}
\end{table}
\end{center}

To test the robustness of our physics-informed approach, we performed 100 independent runs of the GA, each initialized with a different random seed. This procedure ensures that the results are not biased or favored by a particular realization of the random number generator. The left panel of Fig.~\ref{Fig: Seeds} shows the evolution of the fitness for each run across $10^4$ generations. All runs eventually achieve sub-percent fractional error, with several outperforming the zero-baryon EH formula, which reaches $\mape(\EH) = 0.75\%$.

We selected the best-performing seed among these runs and extended its evolution to $10^5$ generations. The right panel of Fig.~\ref{Fig: Seeds} shows that the fitness function plateaus after a few hundred generations, indicating that the GA has reached a stagnation point. The final symbolic expression obtained is:
\begin{equation}
\label{Eq: TGA_nw}
    T_{\GA, \nw}(q) = \left[ 1 + \sum_{i = 1}^{4} a_i q^{b_i}  \right]^{-1/4},
\end{equation}
where the coefficients $\{a_i, b_i\}$ are listed in Table~\ref{Tab: Tnw}. This formula achieves a final fractional error of $\mape(\GA) = 0.67\%$.

Notably, although the grammar allowed for six additive terms, the GA converged on a formula with only four, all of the form $a_i q^{b_i}$, reminiscent of the BBKS structure. The largest exponent in the expression is $b_4 = 7.507$, which implies that, at small scales, the MTS behaves approximately as $T_{\GA, \nw} \propto 1/k^{1.877}$. From physical considerations, one would expect the asymptotic behavior $T(k) \propto \log(k)/k^2$ at small scales~\cite{Zeldovich:1965gev}, which means the GA approximates but does not fully capture the expected scaling. Consequently, we caution against extrapolating this formula beyond the training range, i.e., for $k \gtrsim 1.5~h\,\Mpc^{-1}$. Still, it is worth emphasizing that this upper limit lies well beyond the nonlinear scale, typically around $k_\text{nl} \simeq 0.25~h\,\Mpc^{-1}$~\cite{Dodelson:2020bqr}.

As a baseline for comparison, we also evaluate the performance of a commonly used smoothing technique: the Savitzky-Golay (SG) filter~\cite{Euclid:2023tqw}. This numerical filter is often applied to extract the smooth component of the MTS. We apply it to the same dataset and compute the \mape of the resulting smooth fit. Evaluating all models over the full dataset (i.e., without excluding any points), we obtain:
$$
    \mape(\EH) = 0.93\%,
$$
\begin{equation}
    \mape(\GA) = 0.82\%, \quad \mape(\SG) = 0.79\%.
\end{equation}
The complexities of our formula and the zero-baryon EH formula are measured as:
\begin{align}
    L(\GA) = 62, &\quad D(\GA) = 9, \\
    L(\EH) = 378, &\quad D(\EH) = 20, \nonumber
\end{align}
and thus our formula is around 6 times simpler and around $12\%$ more accurate than the traditional zero-baryon EH formula. Therefore, we conclude that our symbolic expression is a viable analytical alternative, achieving nearly the same level of accuracy as the SG filter while offering the additional benefit of interpretability and analytical differentiability, which can be advantageous in semi-analytical modeling or emulator pipelines.

Finally, we stress again the importance of the training domain in interpreting the results. While our symbolic formula was trained up to $k \sim 1.5~h\,\Mpc^{-1}$, this range extends beyond the linear regime. In practice, linear transfer functions are frequently used as inputs to nonlinear models such as \texttt{halofit} or \texttt{HMcode}, which apply corrections based on fits to N-body simulations~\cite{Smith:2002dz, Takahashi:2012em, Mead:2015yca, Mead:2020vgs}. For instance, the \texttt{Quijote} simulations suite~\cite{Villaescusa-Navarro:2019bje}, with $N_p = 512^3$ particles in a $L = 1000~\Mpc \,h^{-1}$ box, is considered reliable up to the Nyquist mode:
\begin{equation}
    k_\text{Nyq} \equiv \pi \frac{N_p^{1/3}}{L} \simeq 1.6~h\,\Mpc^{-1}.
\end{equation}
Hence, any linear input model—analytic or numerical—used in this regime must be treated with care. For this reason, we will later evaluate our GA-derived expression within \texttt{halofit}, allowing us to probe its performance beyond the training domain.

\subsubsection{\textbf{Test}}

Our main goal is to provide a fully semi-analytical expression for the smooth MPS, $P_{\GA, \nw}(k)$, based on the non-wiggle transfer function. In linear theory, the MPS can be written as:
\begin{equation}
\label{Eq: Pnw}
    P_{\GA, \nw}(k\,|\, \boldsymbol{\theta}) \equiv A_0\, k^{n_s} T^2_{\GA, \nw}(k),
\end{equation}
where $\boldsymbol{\theta} = \{h, \omega_b, \omega_m, n_s, A_s\}$ denotes the cosmological parameters of the standard $\Lambda$CDM model.

The overall amplitude $A_0$ serves as the normalization of the spectrum and depends on these parameters. It is typically fixed by matching the MPS to the observed amplitude of fluctuations on a reference scale, characterized by the variance of the overdensity field smoothed over a radius $R$:
\begin{equation}
\label{Eq: sigma_R}
    \sigma_R^2 \equiv \frac{1}{2\pi^2}\int_0^\infty \dd k\, k^2 P(k)\, W^2(kR),
\end{equation}
where $W(x)$ is the Fourier transform of a real-space top-hat filter:
\begin{equation}
    W(x) \equiv \frac{3}{x^3} \left( \sin x - x \cos x \right).
\end{equation}

Using our expression for the power spectrum, this becomes:
\begin{equation}
\label{Eq: A0}
    \sigma_R^2 = \frac{A_0}{2\pi^2} \int_0^\infty \dd k\, k^{2 + n_s} T^2_{\GA, \nw}(k)\, W^2(kR).
\end{equation}
In practice, we set $R = 8~\Mpc\,h^{-1}$ and infer $A_0$ by matching to the value of $\sigma_8$ computed with \class. Although the definition of $\sigma_8$ formally involves an integral over all wavenumbers, our expression for $T_{\GA,\nw}(k)$ is calibrated only up to $k = 1.5~h\,\Mpc^{-1}$. Consequently, one should avoid relying on contributions from modes beyond this range. However, the top-hat window function provides a sharp ultraviolet suppression, ensuring that modes with $k \gg 1/R$ contribute negligibly to the integral.

To assess the performance of our symbolic formula, we generate a test dataset of 200 cosmologies using a Latin hypercube (LH) sampling over the parameter ranges shown in Table~\ref{Tab: Params}. For each cosmology, we compute the corresponding MPS from \class, and extract the associated value of $\sigma_8$. Note that $\sigma_8$ is not an independent parameter in $\Lambda$CDM, as it is degenerate with $A_s$. In practice, either $\sigma_8$ or $A_s$ can be used as input in \class, or, alternatively, $\sigma_8$ can be computed using accurate fitting formulas such as those in Refs.~\cite{Bartlett:2023cyr, Bartlett:2024jes}. 

Then, we measure the fitness of our reconstructed $P_{\GA, \nw}(k)$ against the output from \class~using the metric defined in Eq.~\eqref{Eq: Acc GA}. On this test dataset, we find:
$$
    \mape(\EH) = 1.68\%,
$$
\begin{equation}
    \mape(\GA) = 0.99\%, \quad \mape(\SG) = 0.99\%,
\end{equation}
which we summarize in Table~\ref{Tab: Tnw metrics}.

\begin{figure}[t!]
\centering    
{\includegraphics[width=\columnwidth]{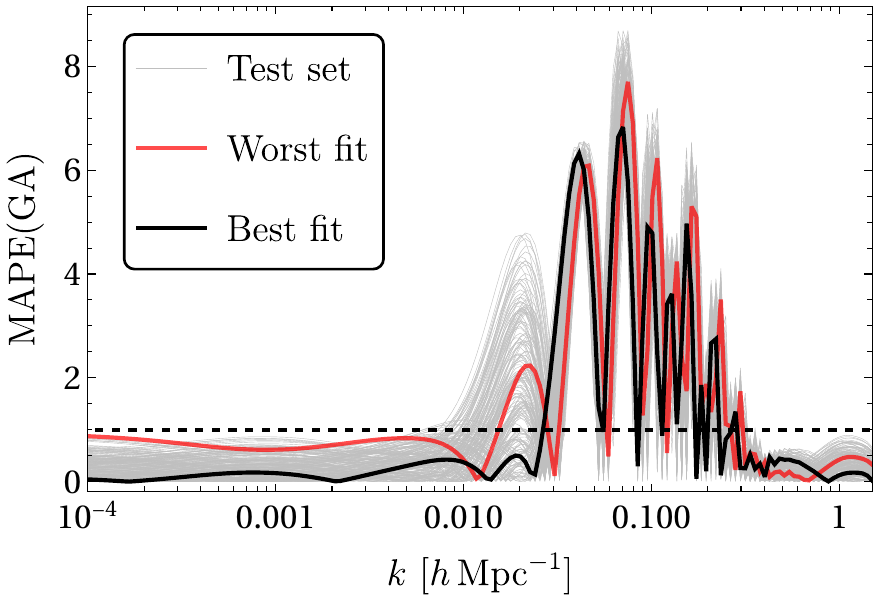} \\[1em]
\includegraphics[width=\columnwidth]{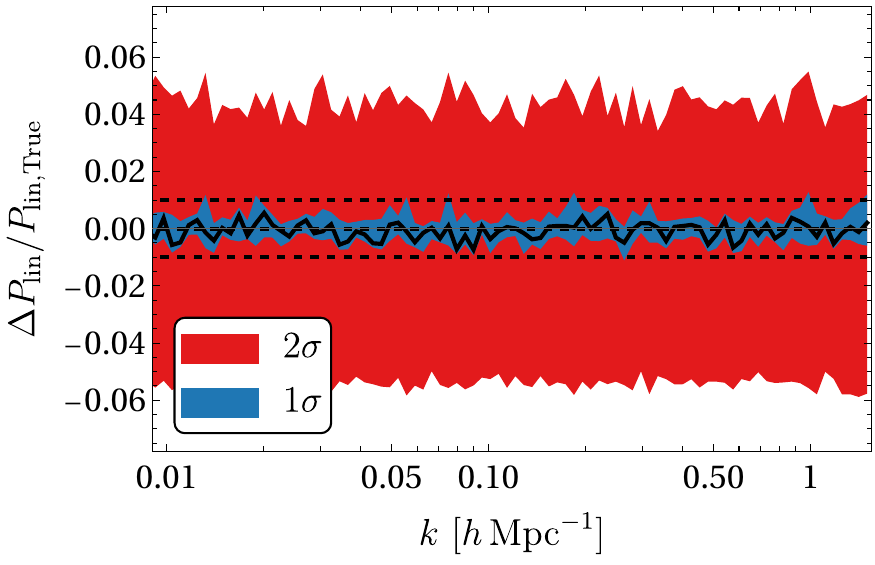}}
\caption{\textbf{Top:} $\mape$ as a function of $k$ for the reconstructed $P_{\GA, \nw}$ across 200 cosmologies sampled in a LH from Table.~\ref{Tab: Params}. Thin gray lines represent individual models; the best and worst cases are highlighted in color. Our formula maintains better than 1\% accuracy across the full range, except for $k \sim 0.01\!-\!0.3~h\,\Mpc^{-1}$, where BAO features dominate. \textbf{Bottom:} Distribution of the fractional errors corresponding to the $1\sigma$ and $2\sigma$ regions. Dashed black lines represent the $1\%$ deviation region.}
\label{Fig: Test Pnw}
\end{figure}

In Fig.~\ref{Fig: Test Pnw}, we show the \mape~as a function of the wavenumber for all 200 cosmologies, and the distribution of the fractional errors defined as:
\begin{equation}
\label{Eq: Fractional diff}
    \frac{\Delta P_\text{lin}}{P_{\text{lin}, \text{True}}} \equiv \frac{P_\text{lin}^\text{(emul)}(k) - P_\text{lin}^\class(k)}{P_\text{lin}^\class(k)},
\end{equation}
considering to the $1\sigma$ and $2\sigma$ deviation regions. The individual accuracy curves are shown as thin gray lines, with the best and worst cases highlighted in color. Our symbolic formula maintains better than 1\% accuracy across all scales, except in the range $k \sim 0.01\!-\!0.3~h\,\Mpc^{-1}$, which coincides with the turnover around matter-radiation equality and the region where BAO features are most prominent. These are precisely the scales where the smooth component should deviate from the full transfer function in order to effectively extract the BAO signal. 

The fact that the error remains small at large scales ($\lesssim 1\%$) also ensures that our formula provides a consistent estimate of the normalization factor $A_0$ via Eq.~\eqref{Eq: A0}. Indeed, we also determined the factor $A_0$ by minimizing the \mape~between the spectra computed using \class~and our GA derived formulation, in order to compare with the factor $A_0$ obtained through the integration explained above. We find that the mean difference between these procedures is about $0.22\%$, providing a direct proof that our calculation of $A_0$ is accurate enough.

Overall, these results confirm that our expression accurately captures the smooth structure of the linear MPS across a wide range of cosmologies, with a precision comparable to that of the SG filter. However, unlike SG and other purely numerical techniques, our symbolic expression yields a compact, differentiable, and interpretable analytical formula. To the best of our knowledge, this is the simplest and most accurate fitting function currently available for the smooth linear MPS. The only comparable analytical alternatives—the BBKS and zero-baryon EH formulas—are either less accurate or more cumbersome within the parameter space considered.

\begin{center}
\begin{table}
\begin{centering}
\begin{tblr}{ c | c | c | c | c }
\hline[1.2pt]
\hline
Model & Mean($\%$) & Median($\%$) & $68\%$ & $95\%$ \\
\hline
\hline[1.2pt]
$P_\nw$ Eq.~\eqref{Eq: Pnw} & 0.99 & 0.99 & 1.06 & 1.21 \\
\hline[1.2pt]
\end{tblr}
\caption{Fractional-error statistics (mean, median, and central credible intervals) characterizing the accuracy of the GA-derived smooth MPS $P_{\nw}$ in Eq.~\eqref{Eq: Pnw}.}
\label{Tab: Tnw metrics}
\end{centering}
\end{table}
\end{center}

\subsection{Wiggles in the Matter Transfer Function}
\label{Sec: Wiggles in the Matter Transfer Function}

\subsubsection{\textbf{Template Function}}

As discussed earlier, once a de-wiggling method is chosen, the BAO signal can be isolated by dividing the full power spectrum, $P(k)$, by its smooth (non-wiggle) counterpart, $P_\nw(k)$. However, the particular de-wiggling strategy adopted can influence the inferred shape of the oscillations~\cite{Chen:2024tfp}. Since acoustic oscillations encode physical information, any de-wiggling process will inevitably transfer modeling assumptions—and potential artifacts—into the extracted BAO signal. These spurious effects are often visible as mismatches at very large or small scales.

Although an exact expression for the wiggles is unattainable, the core physical mechanisms underlying them are well understood. By incorporating this physics into our model, we aim to isolate the acoustic oscillations in a way that minimizes artificial distortions. Below, we outline the key physical ingredients that drive the BAO signal in the MPS.

In Ref.~\cite{Eisenstein:1997ik}, Eisenstein and Hu present a formulation for the total transfer function $T(k)$ that includes all relevant phenomena: acoustic oscillations, Compton drag, velocity overshoot, baryon in-fall, Silk damping, and CDM suppression. Their formula comprises two main contributions: one from CDM and one from baryons. The inclusion of baryons introduces several important effects:
\begin{itemize}
    \item[$i)$] Prior to recombination, baryons are tightly coupled to photons. The acoustic pressure waves imprinted in the CMB also leave their signature in the matter distribution, resulting in the BAO features.
    \item[$ii)$] After decoupling, photon diffusion damps small-scale oscillations—a process known as Silk damping.
    \item[$iii)$] Once free from the photon fluid, baryons fall into CDM potential wells, inducing a further suppression of power at small scales.
\end{itemize}

Our objective is to find an expression for the matter transfer function:
\begin{equation}
    T_\GA(k) \equiv T_{\GA, \nw}(k) T_{\GA, \w}(k),
\end{equation}
such that $T_{\GA, \w}(k)$ encapsulates these effects in a compact and interpretable form optimized via SR with GAs.

We assume that the broadband part of the transfer function, $T_\nw(k)$, is given by the GA-derived expression in Eq.~\eqref{Eq: TGA_nw}. We then model the oscillatory component as follows:
\begin{enumerate}
    \item A sinusoidal modulation capturing the acoustic oscillations inside the sound horizon.
    \item An exponential damping envelope to mimic Silk damping.
    \item An amplitude suppression term reflecting the decreasing strength of oscillations at small scales.
\end{enumerate}
This leads to the following template:
\begin{equation}
\label{Eq: template 1}
    T_{\GA, \w}(k) \equiv \left[1 + f_\text{amp}~e^{-f_\Silk} \sin \left( f_\text{osc} \right) \right].
\end{equation}
The task now becomes finding suitable analytical expressions for the amplitude term $f_\text{amp}(k)$, the damping function $f_\Silk(k)$, and the oscillation phase $f_\text{osc}(k)$, all of which are determined by the GA.\footnote{The symbol $f$ here indicates that these functions are internally optimized by the GA.}

We take inspiration from the EH baryon transfer function in Appendix [Eq.~\eqref{Eq: Tb EH}]. Although this expression is complex, we find that within our parameter range (see Table~\ref{Tab: Params}), several of their constituents can be approximated using simple power-law functions. Based on these considerations, we propose the following templates:
\begin{align}
    f_\text{amp}(k) &\equiv \frac{f_\alpha(\omega_b, \omega_m)}{a_5 + \left\{f_\beta(\omega_b, \omega_m)/(k h s_\GA)\right\}^{b_5}}, \\
    f_\Silk(k) &\equiv a_6(kh/k_\Silk)^{b_6}, \\
    f_\text{osc}(k) &\equiv \frac{a_7 (k h s_\GA + a_8\,\omega_m^{-b_7})}{\left(a_9 + \left\{ f_\text{node}(\omega_m)/(k h s_\GA)\right\}^{b_8} \right)^{b_9}},
\end{align}
where $a_i$ and $b_i$ are free coefficients to be optimized by the GA. The functions $f_\alpha$, $f_\beta$, and $f_\text{node}$ are simple power laws of $\omega_b$ and $\omega_m$, also to be determined by the algorithm.

Instead of using the original EH sound horizon $s_\EH$ [Eq.~\eqref{Eq: sEH}], we adopt the more accurate GA-derived expression from Ref.~\cite{Aizpuru:2021vhd}:
\begin{equation}
    s_\GA \equiv \frac{1}{c_1 \omega_b^{c_2} + c_3\,\omega_m^{c_4} + c_5\omega_b^{c_6}\omega_m^{c_7}},
\end{equation}
with coefficients:
\begin{align}
    c_1 &= 0.00785436, &
    c_2 &= 0.177084, &
    c_3 &= 0.00912388, \nonumber \\
    c_4 &= 0.618711, &
    c_5 &= 11.9611, &
    c_6 &= 2.81343, \nonumber \\
    c_7 &= 0.784719, \label{Eq: Coeff sGA}
\end{align}
which is accurate up to $\sim0.003\%$.

\begin{center}
\begin{table}
\begin{centering}
\begin{tblr}{c c | c c}
\hline[1.2pt]
\hline
Constant & Value & Constant & Value \\
\hline
\hline[1.2pt]
$a_5$ & 1.03922 & $b_5$ & 2.7345 \\
\hline
$a_6$ & 1.36418 & $b_6$ & 0.9914 \\
\hline
$a_7$ & 0.78991 & $b_7$ & 0.26955 \\
\hline
$a_8$ & 0.5426 & $b_8$ & 1.19806 \\
\hline
$a_9$ & 0.7349 & $b_9$ & 0.739668 \\
\hline
$a_{10}$ & 0.10047 & $b_{10}$ & 0.76324 \\
\hline
$a_{11}$ & 0.1485 & $b_{11}$ & 1.49987 \\
\hline
$a_{12}$ & 0.0005236 & $b_{12}$ & 1.47345 \\
\hline
$a_{13}$ & 49.9978 & $b_{13}$ & 0.96164 \\
\hline
$a_{14}$ & 46.02641 & $b_{14}$ & 1.24458 \\
\hline
$a_{15}$ & 1.0049 & $b_{15}$ & 0.234948 \\
\hline[1.2pt]
\end{tblr}
\par\end{centering}
\caption{Coefficients $a_i$ and $b_i$ for the MTS $T_\GA(k)$.}
\label{Tab: Hyperparameters}
\end{table}
\par\end{center}

Similarly, instead of the EH expression for the Silk damping scale $k_\Silk^{(\EH)}$ [Eq.~\eqref{Eq: kSilk EH}], we use the empirically improved formula:
\begin{equation}
    k_\Silk \equiv 0.373\, \omega_b^{0.419} + 0.195\,\omega_m^{1.0957},
\end{equation}
in units of $h\,\Mpc^{-1}$. This formula offers better accuracy in our parameter space ($\mape(k_\Silk) = 0.03\%$). See Appendix~\ref{App: GA formulas} for more details about this expression.

\begin{figure*}[t!]
\centering    
\includegraphics[width=\columnwidth]{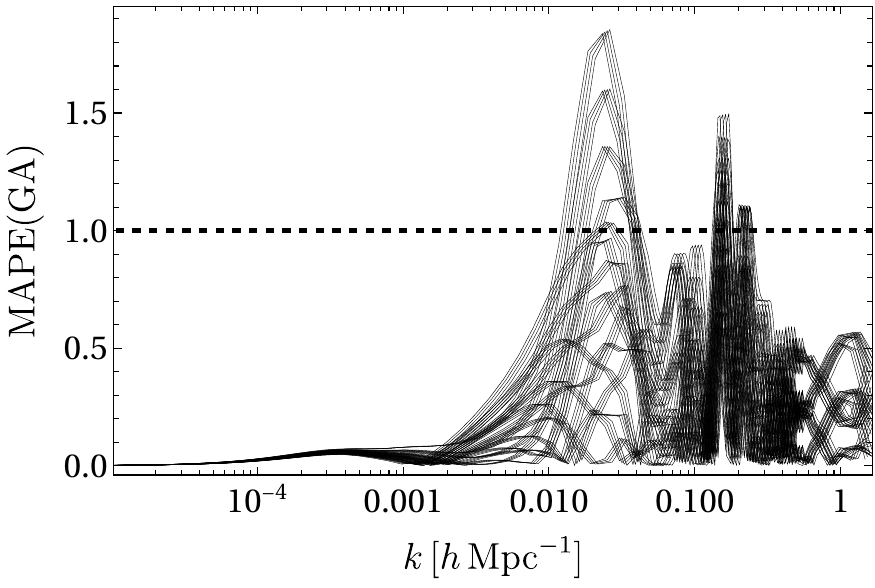} \hfill
\includegraphics[width=\columnwidth]{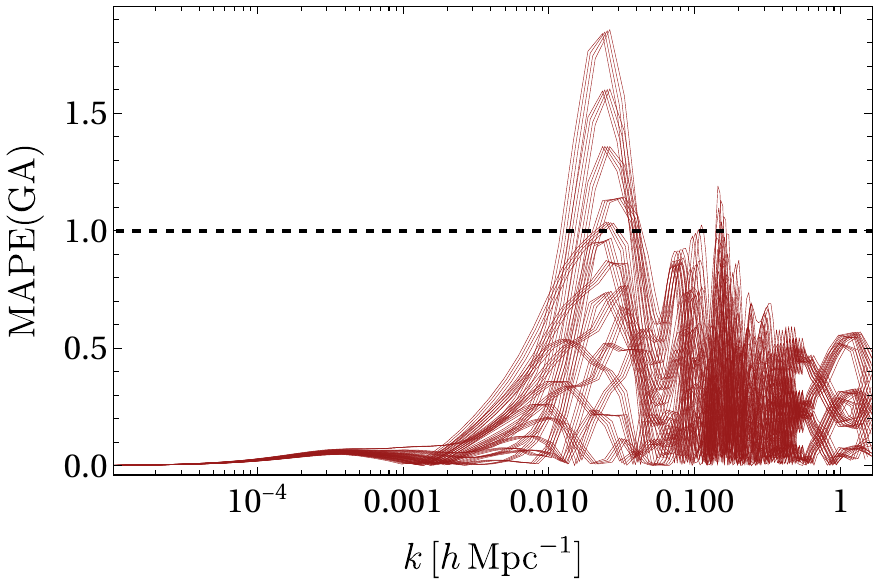}
\caption{Fitness as a function of wavenumber $k$ over the training set. \textbf{Left:} Fitness without correction. Errors peak near $k \sim 0.02~h\,\Mpc^{-1}$ (equality scale) and at $k \sim 0.2~h\,\Mpc^{-1}$ (diffusion scale). The latter feature appears largely independent of the cosmological parameters. \textbf{Right:} Fitness after applying a Gaussian correction around $k \sim 0.2~h\,\Mpc^{-1}$, improving the overall fit to $\mape(\GA) = 0.25\%$.}
\label{Fig: Correction Silk 1}
\end{figure*}

\begin{figure*}
\centering    
{\includegraphics[width=\columnwidth]{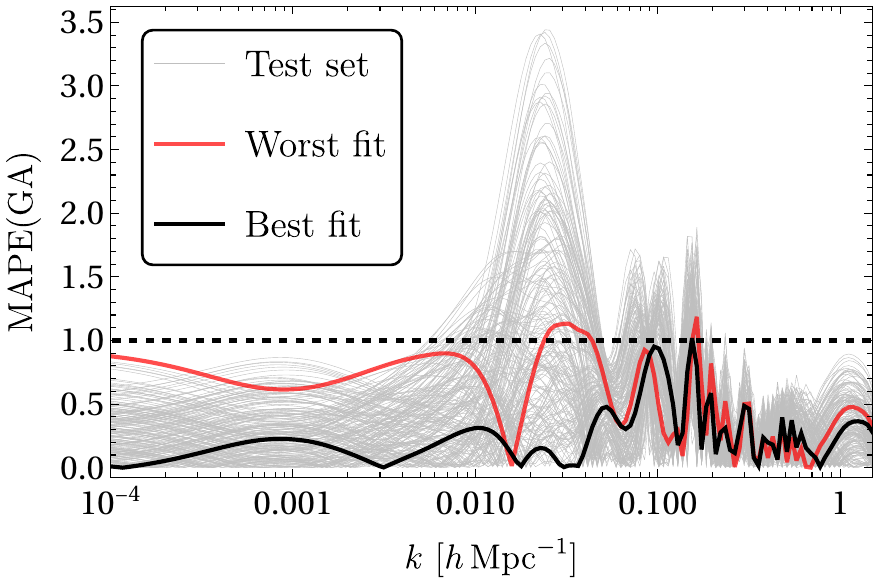} \hfill
\includegraphics[width=\columnwidth]{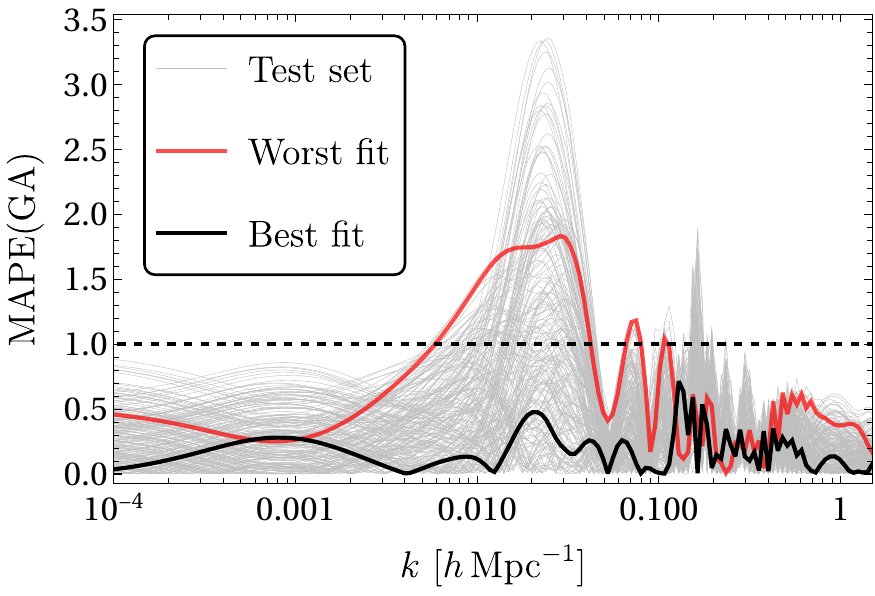} \\[1em]
\includegraphics[width=\columnwidth]{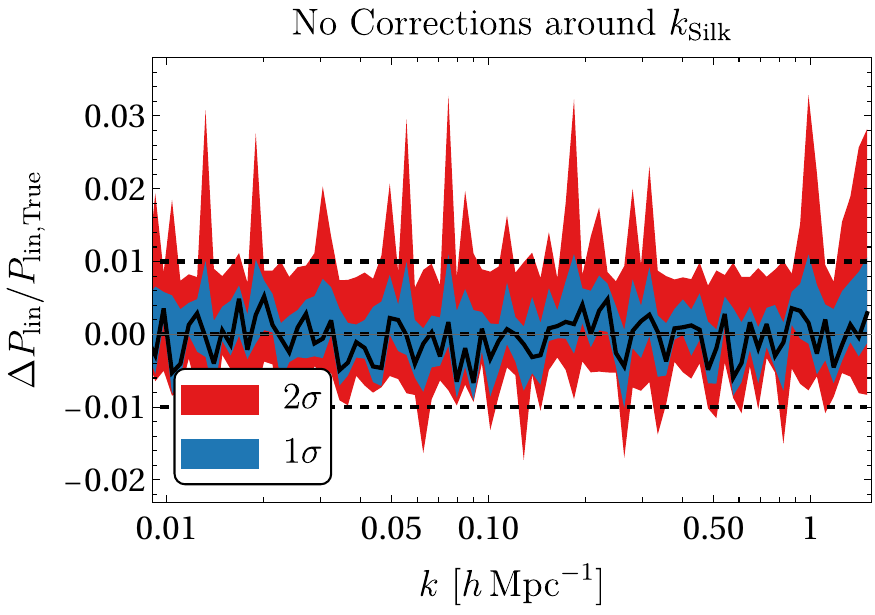} \hfill
\includegraphics[width=\columnwidth]{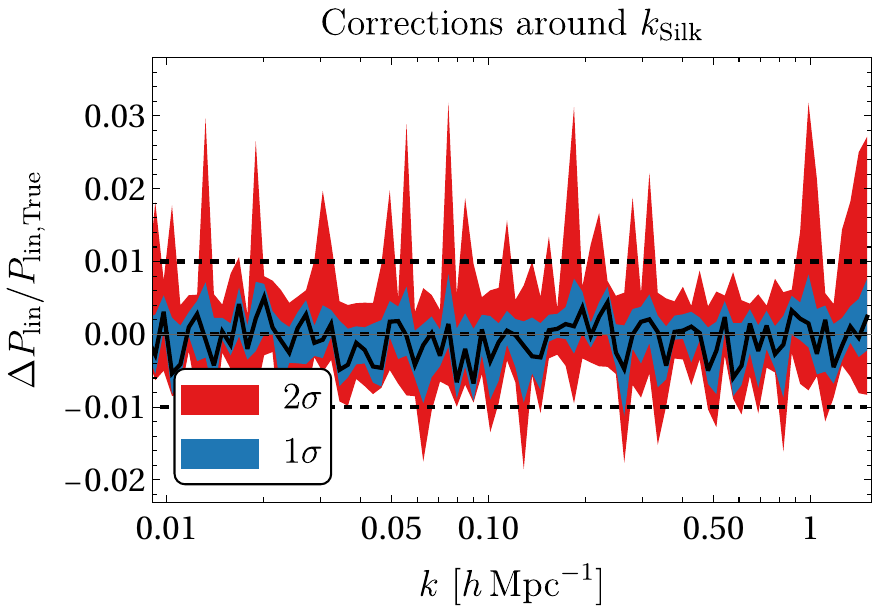}}
\caption{Point-wise fractional error $\mape(k)$ of the emulated linear MPS. \textbf{Top left:} Original model $P_\GA$ achieving $\mape(\GA) = 0.42\%$. Larger errors are observed around $k \sim 0.02~h\,\Mpc^{-1}$ and $k \sim 0.1~h\,\Mpc^{-1}$. \textbf{Top right:} Improved version including three localized Gaussian corrections reduces the mean fractional error to $\mape(\GA) = 0.39\%$, significantly improving the match around the Silk damping scale. Bottom panels show the distribution of the fractional errors corresponding to the $1\sigma$ and $2\sigma$ regions considering no corrections around the Silk scale (\textbf{bottom left}), and considering Gaussian corrections around this scale (\textbf{bottom right}).}
\label{Fig: Test Pk Silk 1}
\end{figure*}

Finally, we reuse the same dataset as in the previous section, without filtering any points—yielding 7296 data samples of the form $\{k, h, \omega_b, \omega_m, T(k)\}$, which is, again, a rather small training dataset. We increase the number of genes and chromosomes in our GA architecture to allow simultaneous optimization of the multiple functional components described above. The performance of each trial is assessed using the same Fitness metric defined in Eq.~\eqref{Eq: Acc GA}.

\subsubsection{\textbf{Fitting Formula}}

We ran the genetic algorithm with 100 different random seeds and selected the best-performing realization for a long run of \(10^5\) generations. The GA then converged to the following functions:
\begin{align}
    f_\alpha (\omega_b, \omega_m) &\equiv a_{10} - a_{11} \omega_b^{b_{10}} + a_{12} \omega_m^{b_{11}}, \\
    f_\beta (\omega_b, \omega_m) &\equiv b_{12} - a_{13} \omega_b^{b_{13}} + a_{14} \omega_m^{b_{14}}, \\
    f_\text{node} (\omega_m) &\equiv a_{15} \omega_m^{b_{15}},
\end{align}
where all constants are listed in Table~\ref{Tab: Hyperparameters}. This solution achieves a training error up to: 
$$
    \mape(\GA) = 0.29\%.
$$
By comparison, the full EH formula yields:
$$
    \mape(\EH) = 0.68\%.
$$
In terms of the complexities, the leaf count and depth of our formula and the full EH formula are measured as:
\begin{align}
    L(\GA) = 214, &\quad D(\GA) = 14, \\
    L(\EH) = 1039, &\quad D(\EH) = 25, \nonumber
\end{align}
and thus our formula is around 5 times simpler and around $60\%$ more accurate than the traditional EH formula.

Figure~\ref{Fig: Correction Silk 1} (left panel) shows the point-wise fitness as a function of $k$ across all training cosmologies. Our formula consistently achieves sub-percent fractional error, with two notable exceptions: $(i)$ a pronounced bump around $k \sim 0.02~h\,\Mpc^{-1}$, which roughly corresponds to the scale of matter–radiation equality $k_\eq$; and $(ii)$ a small bump near $k \sim 0.15{-}0.2~h\,\Mpc^{-1}$, which roughly corresponds to the diffusion scale. Interestingly, this second peak appears to be largely independent of the cosmological parameters.

Upon further analysis, we find that this bias stems from insufficient exponential damping in our model. To mitigate this issue, we introduce a localized Gaussian correction term around the problematic scale:
\begin{equation}
    T_{\S,1}(k) \equiv 1 - A_{\text{S},1} e^{-(k - k_{\text{S},1})^2/\sigma_{\text{S},1}^2},
\end{equation}
with the following parameters (in units of $h\,\Mpc^{-1}$ for $k_{\text{S},1}$ and $\sigma_{\text{S},1}$):
\begin{equation}
\label{Eq: Coeff S1}
    A_{\text{S},1} = -0.00625, \ \ k_{\text{S},1} = 0.199, \ \ \sigma_{\text{S},1} = 0.0627.
\end{equation}
The inclusion of this correction improves the training fitness from $\mape(\GA) = 0.29\%$ to:
$$
    \mape(\GA) = 0.25\%.
$$
This improvement is illustrated in the right panel of Fig.~\ref{Fig: Correction Silk 1}.

\subsubsection{\textbf{Test}}

At this stage, the full MTS is given by:
\begin{equation}
    T_\GA(k) \equiv T_{\GA, \nw}(k)\, T_{\GA, \w}(k)\, T_{\text{S},1}(k),
\end{equation}
and the corresponding linear MPS takes the form:
\begin{equation}
\label{Eq: PGA + Wiggles}
    P_\GA(k) \equiv A_0\, k^{n_s}\, T_\GA^2(k),
\end{equation}
where the normalization factor $A_0$ is computed via Eq.~\eqref{Eq: A0}, using the smooth MTS $T_\nw(k)$ only. This is justified since the normalization—often determined by matching the integrated power to a target $\sigma_8$ value—is sensitive primarily to the broad-band shape, while the oscillatory features contribute as negligible perturbations to the integral in Eq.~\eqref{Eq: sigma_R}.

We reuse the 200 test spectra computed via \class, based on a LH sampling across the parameter ranges in Table~\ref{Tab: Params}. For each cosmology, we retrieve the corresponding value of $\sigma_8$ and use it to fix $A_0$. As shown in the top left panel of Fig.~\ref{Fig: Test Pk Silk 1}, our formula reproduces the numerical MPS with high accuracy at both the largest and smallest scales. The average relative error across the dataset is:
$$
    \mape(\GA) = 0.42\%.
$$

By comparison, the full EH model (as described in Appendix~\ref{App: Full EH}) achieves a lower accuracy:
$$
    \mape(\EH) = 1.63\%,
$$
using the same normalization strategy, with the zero-baryon EH formula as input for $T_\nw$.

Despite this overall good performance, we observe two systematic deviations: a broad peak in the error near $k \sim 0.015{-}0.03~h\,\Mpc^{-1}$, around $k_\eq$, and localized overshooting around $k \sim 0.1~h\,\Mpc^{-1}$, where Silk damping dominates. In both cases, the error can exceed $1\%$, reaching up to $3.5\%$ at the peak near $k_\eq$.

To alleviate the issues around $k_\Silk$, we introduce three additional localized corrections of Gaussian form:
\begin{equation}
    P_{\S,i}(k) \equiv 1 + A_{\S,i}\,e^{-(k - k_{\S,i})^2/\sigma_{S,i}^2},
\end{equation}
where $i = 2, 3, 4$ corresponds to the new corrections centered around the problematic regions. The total corrected power spectrum then becomes:
\begin{equation}
\label{Eq: PGA Silk Corrections}
    P_\GA^\text{(corr)}(k) \equiv P_\GA(k) \times P_{\S,2}(k) \times P_{\S,3}(k) \times P_{\S, 4}(k).
\end{equation}
where the parameters for these corrections, obtained via a simple least-squares optimization of the fitness function, are:
\begin{align*}
    A_{\mathrm{S},2} &= 0.008975, &
    k_{\mathrm{S},2} &= 0.0796, &
    \sigma_{\mathrm{S},2} &= 0.03616, \\
    A_{\mathrm{S},3} &= -0.010198, &
    k_{\mathrm{S},3} &= 0.14326, &
    \sigma_{\mathrm{S},3} &= 0.00849, \\
    A_{\mathrm{S},4} &= -0.003792, &
    k_{\mathrm{S},4} &= 1.2011, &
    \sigma_{\mathrm{S},4} &= 0.31853,
\end{align*}
with $k_{\S, i}$ and $\sigma_{\S, i}$ in units of $h\,\Mpc^{-1}$. 

\begin{table*}
\begin{centering}
\begin{tblr}{ c | c | c | c | c }
\hline[1.2pt]
\hline
Model & Mean($\%$) & Median($\%$) & $68\%$ & $95\%$ \\
\hline
\hline[1.2pt]
$P_\GA$ Eq.~\eqref{Eq: PGA + Wiggles} & 0.42 & 0.43 & 0.48 & 0.62 \\
\hline
$P_\GA$ $+$ Gaussians around $k_\Silk$ Eq.~\eqref{Eq: PGA Silk Corrections} & 0.38 & 0.38 & 0.44 & 0.57 \\
\hline[1.2pt]
\end{tblr}
\caption{Fractional-error statistics (mean, median, and central credible intervals) characterizing the accuracy of the GA-derived MPS $P_\GA$ in Eq.~\eqref{Eq: PGA + Wiggles}, and the MPS $P^{(\text{corr})}_\GA$ in Eq.~\eqref{Eq: PGA Silk Corrections} considering Gaussian corrections around $k_\Silk$.}
\label{Tab: PGA metrics kSilk}
\end{centering}
\end{table*}

After applying these corrections, the accuracy improves to:
$$
    \mape(\GA) = 0.39\%.
$$

As shown in the top right panel of Fig.~\ref{Fig: Test Pk Silk 1}, these Gaussian factors effectively suppress the largest residuals in the regions dominated by photon diffusion. Most spectra in the test set now remain within 1\% fractional error across all relevant scales. The corresponding error distributions, displayed in the bottom panels of Fig.~\ref{Fig: Test Pk Silk 1}, further demonstrate the significant improvement over the original non-wiggle MPS, highlighting the effectiveness of the correction. A summary of the associated fitness metrics is provided in Table~\ref{Tab: PGA metrics kSilk}.

\subsubsection{\textbf{Corrections around $k_\eq$}}

\begin{figure}[t!]
\centering    
\includegraphics[width=\columnwidth]{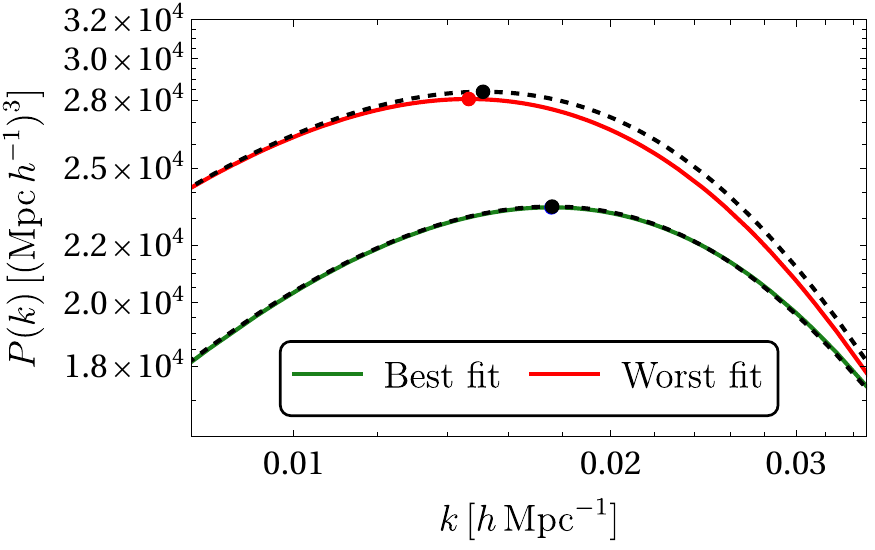}
\caption{Best-fit (green) and worst-fit (red) examples from the 200 test cosmologies. The worst case fails to match the location and amplitude of the peak at $k_\max$, as indicated by the \class~prediction (black dashed line).}
\label{Fig: Problem keq}
\end{figure}

The error peak near $k_\eq$ exhibits strong cosmology dependence, unlike the more systematic error around $k_\Silk$. To diagnose this issue, we compare the best and worst fits in the test set. As shown in Fig.~\ref{Fig: Problem keq}, the best-fit case accurately reproduces both the position and amplitude of the peak in $P(k)$, while the worst-fit spectrum fails to do so.

To address this, we introduce a smooth correction modeled as a skew-normal function centered at $k_\max$, i.e., where the maximum of $P(k)$ is located:
\begin{align}
    &P_\max(k) \equiv 1 \nonumber \\
    &+ \left\{A_\max(k - k_\max) + B_\max\right\} 
    e^{-\frac{1}{2}(k - k_\max)^2/\sigma_\max^2} \nonumber \\
    &\times \left[1 + \text{Erf}\left(\lambda_\max \frac{k - k_\max}{\sqrt{2}\sigma_\max}\right) \right], 
\end{align}
and thus the total MPS will be given by:
\begin{equation}
\label{Eq: Total GA MPS}
    P_\text{full}(k) \equiv P_\GA^{(\text{corr})}(k) \times P_\max(k)
\end{equation}
where $A_\max$, $B_\max$, $\lambda_\max$, and $\sigma_\max$ are parameters to be determined for each cosmology. Here, the variability across cosmologies requires these parameters to depend explicitly on the cosmological inputs.

\vspace{1em}
\noindent\textbf{Amplitude Correction.} The parameter $B_\max$ is fixed by requiring that the corrected spectrum matches the \class~output at the peak:
\begin{equation}
    P_\class(k_\max) = P_\GA(k_\max)\, P_\max \quad \Rightarrow
\end{equation}
\begin{equation}
    B_\max = R_\max - 1, \quad R_\max \equiv \frac{P_\class(k_\max)}{P_\GA(k_\max)}.
\end{equation}
Empirically, we find that the following fitting function accurately captures the behavior of $R_\max$ (up to 0.085\% error):
\begin{align}
    R_\max &= 0.6461 - 0.0097 h^2 + 0.0307 n_s \nonumber \\
          &\quad + 7.1728 \omega_b + 0.0239\, \omega_m^{-1}.
\end{align}

\noindent\textbf{Peak Location Correction.} The parameter $A_\max$ is determined from the condition that the corrected spectrum has zero derivative at $k_\max$:
\begin{equation}
    \left.\frac{\dd}{\dd k} \Big( P_\GA(k)\, P_\max(k) \Big) \right|_{k_\max} = 0.
\end{equation}
This yields:\footnote{Here, a prime denotes the derivative with respect to $k$.}
\begin{align}
    A_\max = &-\left( \frac{P'_\GA(k_\max)}{P_\GA(k_\max)} \right)(1 + B_\max) \nonumber \\
    &- \sqrt{\frac{2}{\pi}} \left( \frac{\lambda_\max}{\sigma_\max} \right) B_\max.
\end{align}

To avoid evaluating derivatives of the full expression, we provide fitting formulas for \(P_\GA(k_\max)\) and its derivative:
\begin{align}
    \frac{P_\GA(k_\max)}{A_0} &= 0.047\, \frac{\omega_m^{1.06224}\, e^{-8.50419\omega_b}}{h^{0.89981}\, n_s^{3.91548}}, \\
    \frac{P'_\GA(k_\max)}{A_0} &= -0.40649 + 0.00518\, h - 0.00139\, h^2 \nonumber \\
    &\quad + 0.26388\, n_s - 0.10340\, n_s^2 \nonumber \\
    &\quad - 2.22729\, \omega_b - 18.4715\, \omega_b^2 \nonumber \\
    &\quad + 3.0768\, \omega_m - 6.70755\, \omega_m^2,
\end{align}
accurate to 0.055\% and 0.04\%, respectively.

\vspace{1em}
\noindent\textbf{Width and Asymmetry Corrections.} 
The curvature of the peak is matched by equating second derivatives:
\begin{equation}
    \left. \frac{\dd^2}{\dd k^2} P_\class\right|_{k_\max} 
    = \left. \frac{\dd^2}{\dd k^2} \left( P_\GA \, P_\max \right) \right|_{k_\max}.
\end{equation}
For fixed $\lambda_\max$, we observe that the second derivative has the following functional form:
\begin{equation}
    \left. \frac{\dd^2}{\dd k^2} (P_\GA \, P_\max) \right|_{k_\max} = d_0 + d_1\, \sigma_\max^{-1} + d_2\, \sigma_\max^{-2},
\end{equation}
with coefficients $d_i$ determined from cosmology.

\begin{table*}
\begin{centering}
\begin{tblr}{ c | c | c | c | c }
\hline[1.2pt]
\hline
Model & Mean($\%$) & Median($\%$) & $68\%$ & $95\%$ \\
\hline
\hline[1.2pt]
No-Wiggly MPS $P_\nw$ Eq.~\eqref{Eq: Pnw} & 0.99 & 0.99 & 1.06 & 1.21 \\
\hline
Wiggly $P_\GA$ Eq.~\eqref{Eq: PGA + Wiggles} & 0.42 & 0.43 & 0.48 & 0.62 \\
\hline
$P_\GA$ $+$ Gaussians around $k_\Silk$ Eq.~\eqref{Eq: PGA Silk Corrections} & 0.39 & 0.40 & 0.44 & 0.63 \\
\hline
$P_\GA$ $+$ corrections around $k_\Silk$ and $k_\eq$ in Eq.~\eqref{Eq: Total GA MPS} & 0.28 & 0.27 & 0.30 & 0.44 \\
\hline[1.2pt]
\end{tblr}
\caption{Fractional-error statistics (mean, median, and central credible intervals) characterizing the accuracy of all GA-derived MPS in this work.}
\label{Tab: PGA full metrics}
\end{centering}
\end{table*}

\begin{figure}[t!]
\centering    
\includegraphics[width=\columnwidth]{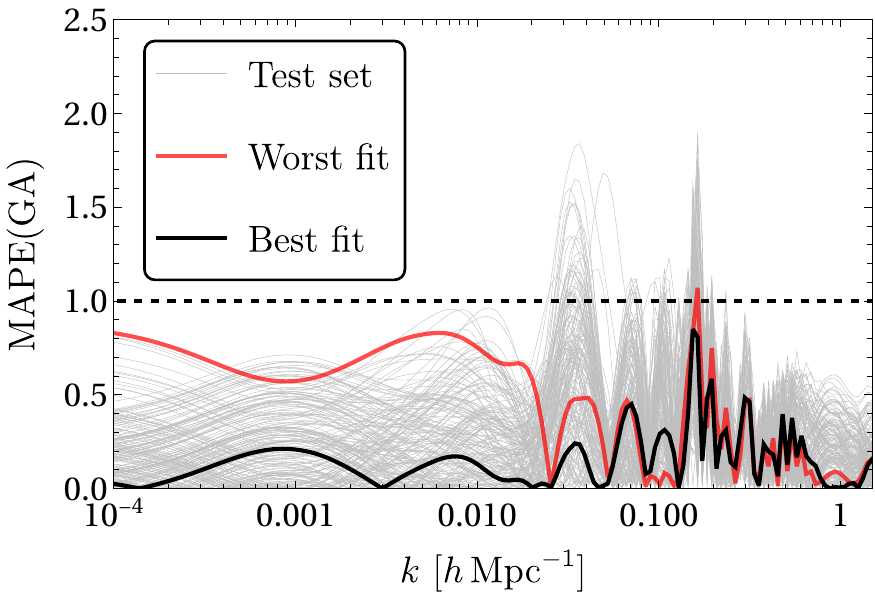}
\caption{Fitness as a function of $k$ for the fully corrected linear MPS across 200 cosmologies. The worst deviations occur at very large scales due to normalization, which, however, are below 1\%. Most models remain below 1\% error across all scales. Final mean fractional error: $\mape(\GA) = 0.28\%$}.
\label{Fig: Full correction}
\end{figure}

\begin{figure*}[t!]
\centering    
\includegraphics[width=\columnwidth]{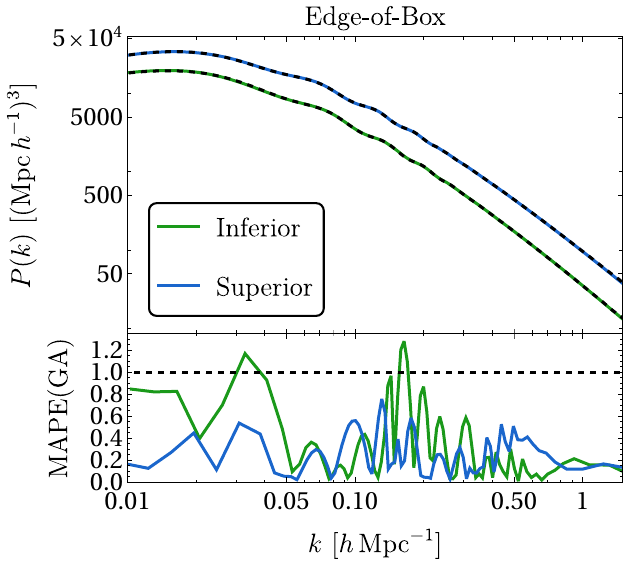} \hfill
\includegraphics[width=\columnwidth]{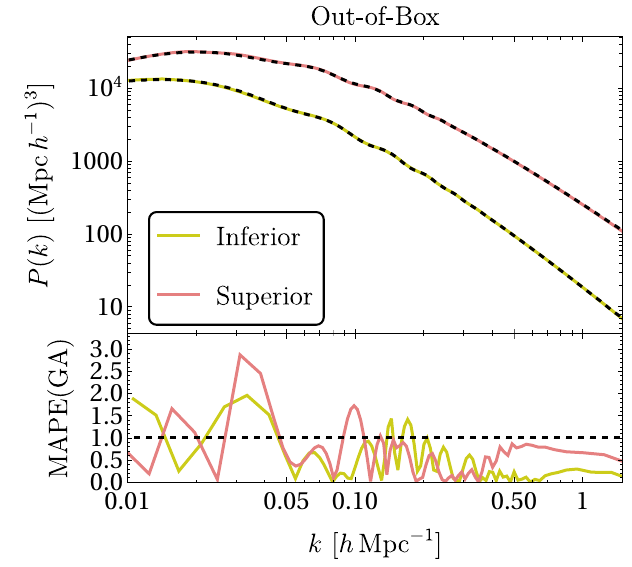}
\caption{The performance of our emulator is illustrated for two edge-of-box parameter choices (\textbf{left}) and two out-of-box realizations (\textbf{right}). The top panels show the MPS predicted by the emulator (solid lines) compared with the numerical results from \class~(dashed black lines). The bottom panels show the corresponding fitness as a function of $k$. These examples demonstrate that the emulator not only performs accurately near the boundaries of the training domain, but also generalizes robustly beyond it.}
\label{Fig: Edge and Out}
\end{figure*}

Since we did not find any informative prior yielding to an expression for $P''_\class(k_\max)$, we used symbolic regression via \texttt{PySR} for this quantity (see Footnote~\ref{FN: GA code}), obtaining:
\begin{align}
   \log_{10} \left[ P''_\class(k_\max) \right] &= 
   10.9041 - \cos h - \frac{1.1084}{h + 3\omega_b} + \omega_b \nonumber \\
   &+ \log\left( \frac{(A_s/10^9)^{0.4338}}{n_s} \right),
\end{align}
with a mean squared error (MSE) of \(7.27 \times 10^{-7}\). Solving the resulting quadratic equation yields two possible values for \(\sigma_\max\); empirically, the smaller root provides the best correction. We also fix \(\lambda_\max \sim 0.4\), implying a slower decay on the high-\(k\) side of the peak—consistent with the observed asymmetry.

\vspace{1em}
\noindent\textbf{Location of the Maximum.}
Finally, we use the following accurate expression (0.044\% error) for \(k_\max\) in units of $[h\,\Mpc^{-1}]$:
\begin{equation}
    k_\max = \frac{0.07066\,\omega_m^{0.8824}\, n_s^{0.939}}{h^{1.00649}(1 + 1.2025\, \omega_b)^{3.3395}}.
\end{equation}
For derivation details, see Appendix~\ref{App: GA formulas}.

\begin{figure*}[t!]
\centering    
\includegraphics[width=0.46\textwidth]{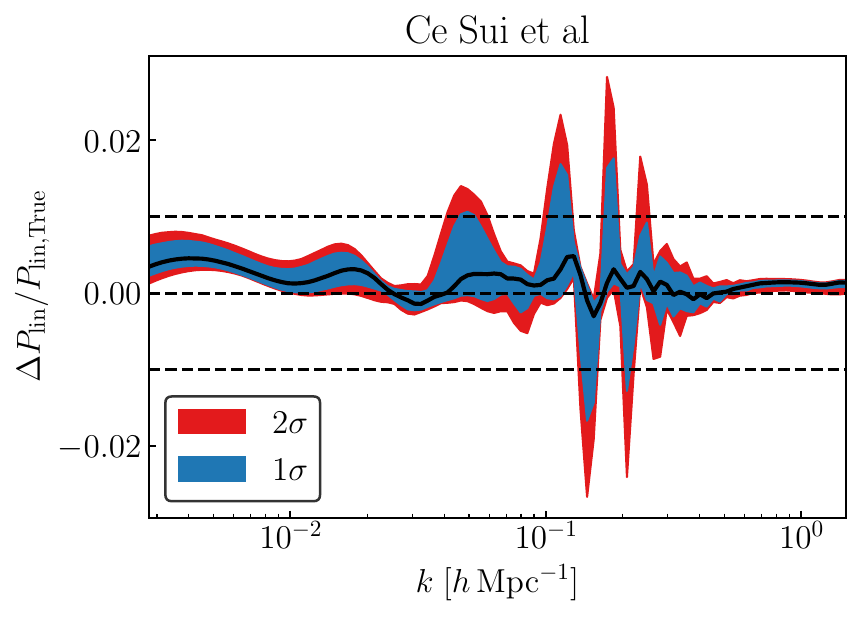} \hfill
\includegraphics[width=0.48\textwidth]{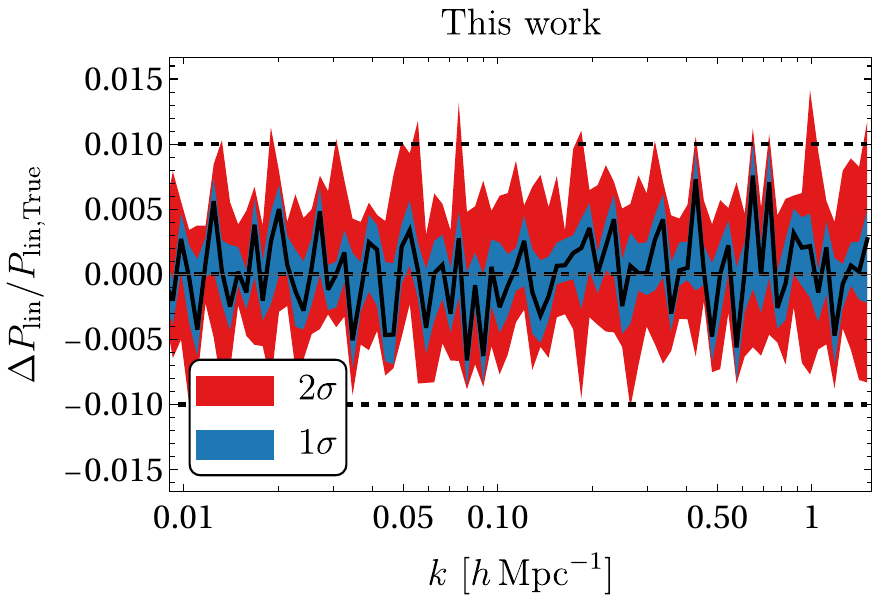}
\caption{Distribution of the fractional errors of linear $P(k)$ emulators from Ref.~\cite{Sui:2024wob} (\textbf{left}) and from this work (\textbf{right}). The shaded bands correspond to the $1\sigma$ and $2\sigma$ regions computed across a Latin hypercube suitable for both emulators. Dashed lines indicate the 1\% error level. When averaged over $k$, our formula achieves a mean fractional error of $\mape(\GA) = 0.28\%$, while that of Sui et al. reaches $\mape(\text{Sui}) = 0.24\%$}.
\label{Fig: Band Plot}
\end{figure*}

\vspace{1em}
\noindent\textbf{Final Accuracy and Complexity.}
With all corrections applied, we re-evaluate the full model against the 200 test spectra. As shown in Fig.~\ref{Fig: Full correction}, nearly all predictions now lie within 1\% of the \class~output. The remaining errors stem primarily from inaccuracies in the amplitude normalization $A_0$ at very large scales, where the weight in the error is highest but below $1\%$ in all instances.

The final average error achieved is:
\begin{equation}
    \mape(\GA) = 0.28\%.
\end{equation}
The symbolic complexity of the full model is:
\begin{equation}
    L(\GA) = 482, \quad D(\GA) = 15,
\end{equation}
with the only external steps being the evaluation of an integral for $A_0$ and solving a quadratic equation for $\sigma_\max$. Recalling than the full EH formula achieves $\mape(\EH) = 1.63\%$ on this test set, we conclude that our full emulator is around 4 times simpler and $82\%$ more accurate than the traditional full EH formulation. 

In Table~\ref{Tab: PGA full metrics}, we summarize the accuracy metrics for all the proposals: no-wiggles MPS in Eq.~\eqref{Eq: Pnw}, MPS with wiggles in Eq.~\eqref{Eq: PGA + Wiggles}, MPS with wiggles and corrections around $k_\Silk$ in Eq.~\eqref{Eq: PGA Silk Corrections}, and the MPS considering corrections around $k_\eq$ in Eq.~\eqref{Eq: Total GA MPS}.

For completeness, in Fig.~\ref{Fig: Edge and Out}, we show two edge-of-box and two out-of-box emulated MPS using our full formulation. For the edge-of-box realizations we take the minimum and maximum values in the training ranges to emulate the MPS plotted as ``inferior'' and ``superior'', respectively. For the out-of-box realizations, we take parameters out of these ranges. For reference, the parameter vectors are reported in Tab.~\ref{tab:vactor_params}.

\begin{table}[h]
\centering
\caption{The parameter vectors used in this analysis.}
\label{tab:vactor_params}

\resizebox{\columnwidth}{!}{%
\begin{tblr}{l | c | c | c | c | c | c }
\hline[1.2pt]
\hline
        & $h$  & $\omega_b$ & $\omega_m$ & $n_s$ & $10^9A_s$ & MAPE \\
\hline[1.2pt]
\hline
Edge-inf & 0.65 & 0.0214    & 0.13       & 0.9   & 1.5       & 0.34\% \\
\hline
Edge-sup & 0.75 & 0.0234    & 0.15       & 1.0   & 2.5       & 0.26\% \\
\hline
Out-inf: & 0.61 & 0.02 & 0.13 & 0.85 & 1.0 & 0.49\%\\
\hline
Edge-inf: & 0.8 & 0.025 & 0.175 & 1.2 & 2.8 & 0.67\%\\
\hline[1.2pt]
\end{tblr}%
}
\end{table}


\subsection{Comparison with Other Formulations}
\label{Sec: Comparison}

\subsubsection{\textbf{Accuracy}}

As mentioned in the introductory section, an alternative SR formulation for the linear MPS was proposed by Bartlett et al. in Ref.~\cite{Bartlett:2023cyr} and subsequently refined by Sui et al. in Ref.~\cite{Sui:2024wob}. Their resulting expression represents an excellent fit to numerical results, reporting a root-mean-square (RMS) fractional error of approximately 0.4\%. Here, we perform a quantitative comparison between their formulation and ours, using their publicly available Python package \texttt{symbolic\_pofk}.\footnote{\href{https://github.com/DeaglanBartlett/symbolic_pofk}{https://github.com/DeaglanBartlett/symbolic\_pofk}}

To this end, we generate 200 test spectra using \class~over a Latin hypercube designed to be valid for both emulators. For instance, while the formulation in Ref.~\cite{Sui:2024wob} is valid over the range $k \in [9 \times 10^{-3}, 9]~h\,\Mpc^{-1}$, our formulation is trained over $k \in [10^{-5}, 1.5]~h\Mpc^{-1}$. Therefore, to ensure a fair comparison, we restrict our evaluation to the overlapping domain $k \in [9 \times 10^{-3}, 1.5]~h\,\Mpc^{-1}$.

We compute the fractional error as defined in Eq.~\eqref{Eq: Fractional diff} and summarize the results in Fig.~\ref{Fig: Band Plot}. Applying our fitness metric [Eq.~\eqref{Eq: Acc GA}], we find:
\begin{equation}
    \mape(\text{Sui}) = 0.24\%, \quad \mape(\GA) = 0.28\%.
\end{equation}

\begin{figure}[t!]
\centering    
\includegraphics[width=\columnwidth]{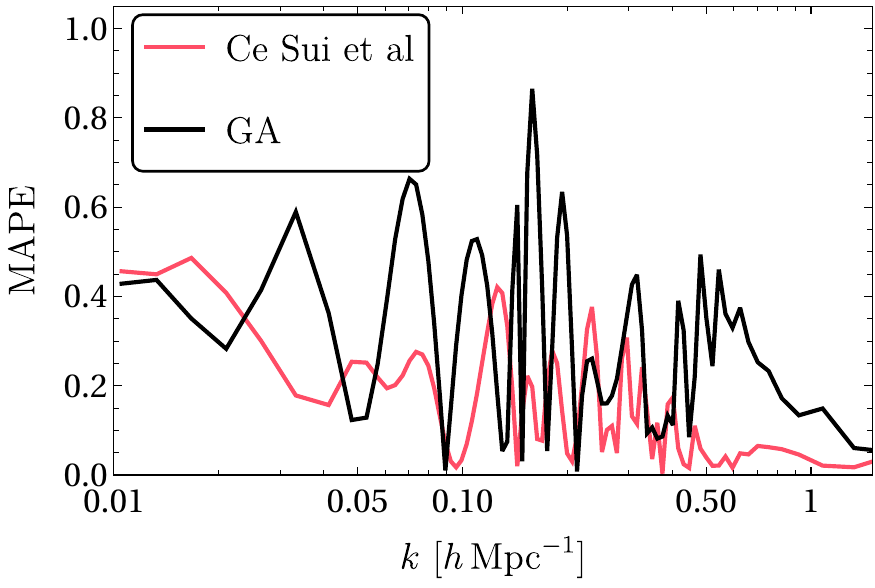}
\caption{Fitness as a function of $k$ for the fully corrected linear MPS in the \texttt{fiducial} cosmology used in the \texttt{Quijote} simulations. Both our formulation and the \texttt{symbolic\_pofk} model remain below 1\% error across all scales.}
\label{Fig: Fiducial}
\end{figure}

We also evaluate both formulations on a specific cosmological model: the \texttt{fiducial} cosmology from the \texttt{Quijote} simulations~\cite{Villaescusa-Navarro:2019bje}, defined by:
\begin{itemize}
    \item[] $h = 0.6711$, \quad $\Omega_b = 0.049$, \quad $\Omega_m = 0.3175$,
    \item[] $n_s = 0.9624$, \quad $\sigma_8 = 0.834$.
\end{itemize}
From these parameters, we compute $\omega_b$, $\omega_m$, and $A_s$ to feed into our formulation. The fitness as a function of $k$ for this cosmology is shown in Fig.~\ref{Fig: Fiducial}. We obtain $\mape(\EH) = 2.19\%$ and:
\begin{equation}
    \mape(\text{Sui}) = 0.16\%, \quad \mape(\GA) = 0.32\%.
\end{equation}

The complexity of the symbolic formula from Ref.~\cite{Sui:2024wob} is measured as:
\begin{equation}
    L(\text{Sui}) = 616, \quad D(\text{Sui}) = 20,
\end{equation}
which is higher than the complexity of our GA-derived model. Although their formulation is more accurate, we emphasize additional criteria that are equally important when assessing SR models intended for physical observables.

In particular, while the formula in \texttt{symbolic\_pofk} achieves excellent numerical performance, its structure is significantly more opaque. Their wiggle component is introduced on top of the zero-baryon EH model, multiplied by a complex correction factor. This oscillatory correction takes the form:
\begin{equation}
\label{eq:lcdm_linear_F}
\begin{split}
    \log F &=  b_0 h - b_{1} + \left( \frac{b_2 \Omega_{\rm b}}{\sqrt{h^2 + b_3}} \right)^{b_4 \Omega_{\rm m}} \\
    &\quad \times \Biggl[
        \frac{(b_5 k - \Omega_{\rm b})b_8}{\sqrt{b_6 + (\Omega_{\rm b} - b_7 k)^2}} (b_9 k)^{-b_{10} k} \\
        &\qquad \times \cos \left( b_{11} \Omega_{\rm m} - \frac{b_{12} k}{\sqrt{b_{13} + \Omega_{\rm b}^2}} \right) \\ 
        &\qquad - b_{14} \left( \frac{b_{15} k}{\sqrt{1 + b_{16} k^2}} - \Omega_{\rm m} \right) \\
        &\qquad \times \cos \left( \frac{b_{17} h}{\sqrt{1 + b_{18} k^2}} \right)
    \Biggr] \\
    &\quad + b_{19} (b_{20} \Omega_{\rm m} + b_{21} h - \log(b_{22} k) + (b_{23} k)^{- b_{24} k}) \\
    &\qquad \times \cos \left( \frac{b_{25}}{\sqrt{1 + b_{26} k^2}} \right) \\
    &\quad + (b_{27} k)^{-b_{28} k} \left( b_{29} k - \frac{b_{30} \log(b_{31} k)}{\sqrt{b_{32} + (\Omega_{\rm m} - b_{33} h)^2}} \right) \\
    &\qquad \times \cos \left(  b_{34} \Omega_{\rm m} - \frac{b_{35} k}{\sqrt{b_{36} + \Omega_{\rm b}^2}} \right),
\end{split}
\end{equation}
where $b_i$ are parameters properly given in Ref.~\cite{Bartlett:2023cyr}.

Although some terms—such as those involving cosine modulations and factors like $1/\sqrt{b_{36} + \Omega_{\rm b}}$—can be heuristically associated with the comoving sound horizon scale, other elements such as $(b_{20} \Omega_{\rm m} + b_{21} h - \log(b_{22} k) + (b_{23} k)^{- b_{24} k})$ are considerably harder to link to known physical processes.

In contrast, our GA-based formulation remains largely interpretable, with the exception of the correction terms introduced around $k_\Silk$ and $k_\eq$, which are not meant to model additional physical mechanisms but rather to finely adjust residual discrepancies between the template and the numerical data. All other components of our template model are grounded in well-understood physical effects. For instance, we explicitly incorporate the expected exponential damping caused by photon diffusion at the Silk scale and employ a simple yet accurate expression to determine the corresponding wavenumber $k_\Silk$.

Therefore, while less accurate, our formulation offers the significant advantage of physical transparency, enabling robust interpretation, extension, and potential calibration with additional theoretical inputs. Importantly, the resulting sub-percent level accuracy,
$$
\Acc(\GA) < 1\%,
$$
fulfills the precision requirements for modern cosmological analyses~\cite{Taylor:2018nrc}.

\subsubsection{\textbf{Runtime}}

To evaluate the computational performance of our emulator, we compute the linear MPS for a Latin Hypercube sample of 1000 realizations drawn from the parameter ranges listed in Table~\ref{Tab: Params}. The runtime benchmarks were executed on a workstation with an AMD Ryzen Threadripper PRO 7975WX CPU (32 cores / 64 threads). Our emulator is implemented in \texttt{Mathematica}, the formulas of Refs.~\cite{Bartlett:2023cyr, Bartlett:2024jes, Sui:2024wob} are evaluated using the \texttt{symbolic\_pofk} Python package, and the baseline linear spectra are generated with the Python wrapper of \class.

The resulting execution times are presented in Table~\ref{Tab: Runtime}. Our interpretable emulator, $P_\GA$, requires approximately $150$ ms to evaluate 1000 spectra, making it about 500 times faster than \class~and comparable in speed to \texttt{symbolic\_pofk}. The inclusion of localized corrections increases the total runtime by roughly a factor of three. These corrections are currently incorporated solely to enhance accuracy; however, they also suggest a clear path toward further improving our formula by introducing additional physical priors. For example, one could replace the integral expression for $A_0$ in Eq.~\eqref{Eq: A0} with an explicit function of the cosmological parameters, following the approach used in \texttt{symbolic\_pofk}, which achieves part of its speed from this substitution. Exploring such refinements is a promising direction for future work.

\begin{center}
\begin{table}
\begin{centering}
\begin{tblr}{ c | c }
\hline[1.2pt]
\hline
Evaluation & time (ms) \\
\hline
\hline[1.2pt]
$A_0$ & 66 \\
\hline
Corrections to $P_\GA$ & 336 \\
\hline
$T_\nw$ & 18 \\
\hline
$T_\GA$ & 56 \\
\hline
$A_0$ and $P_\GA$ & $150$ \\
\hline
$A_0$ and $P_\GA+$corrections & 522 \\
\hline
\texttt{symbolic\_pofk} & 125 \\
\hline
\class & $7.1 \times 10^4$ \\
\hline[1.2pt]
\end{tblr}
\caption{Total runtime spent by our emulators, \texttt{symbolic\_pofk}, and \class, evaluating 1000 spectra for a LH cube from Table~\ref{Tab: Params}.}
\label{Tab: Runtime}
\end{centering}
\end{table}
\end{center}

\subsection{Extension to Nonlinear Scales}
\label{Sec: NL Extension}

Our emulator is trained to reproduce the linear MPS with high accuracy down to $k = 1.5 \, h\,\Mpc^{-1}$. At these scales, however, linear theory no longer provides a reliable description of structure growth, and nonlinear (NL) effects must be taken into account. These contributions can be modeled using semi-analytical prescriptions such as \texttt{halofit} or \texttt{HMcode}~\cite{Smith:2002dz, Takahashi:2012em, Mead:2015yca, Mead:2020vgs, Zhao:2013dza, Gupta:2023rbf, Bose:2020wch}, which apply NL corrections to a given linear MPS. In our case, we feed \texttt{halofit} with the linear MPS predicted by our semi-analytical emulator and compare the resulting nonlinear spectrum with the fully numerical \texttt{halofit} output obtained directly from \class.

\begin{figure}[t!]
\centering    
\includegraphics[width=\columnwidth]{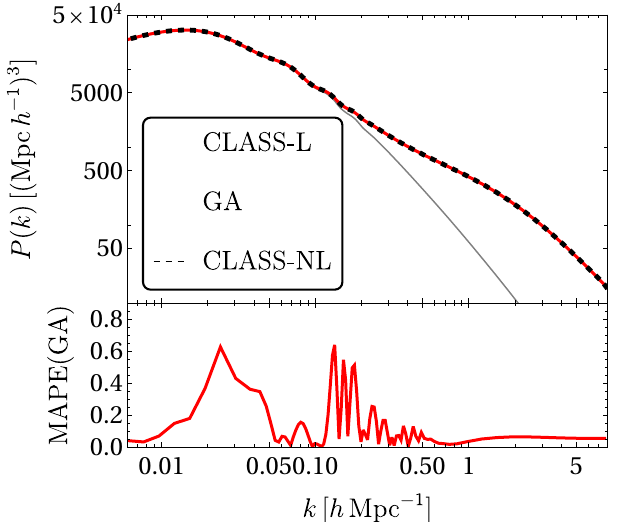}
\caption{\textbf{Top:} Linear (gray) and nonlinear (black dashed) MPS computed with \class~for a representative cosmology, compared against our emulator prediction (red), which is used as input to \texttt{halofit}. \textbf{Bottom:} Fractional accuracy as a function of $k$. In this example,
the error remains below $1\%$ up to $k \sim 8\,h\,\Mpc^{-1}$.}
\label{Fig: Halofit}
\end{figure}

The dimensionless NL MPS, using the revised \texttt{halofit} formula of Takahashi et al.~\cite{Takahashi:2012em}, is decomposed as:
\begin{equation}
    \Delta^2(k) \equiv \frac{k^3 P(k)}{2\pi^2}
    = \Delta_{\mathrm{Q}}^2(k) + \Delta_{\mathrm{H}}^2(k),
\end{equation}
where $\Delta_{\mathrm{Q}}^2$ is the quasi-linear (two-halo) term and $\Delta_{\mathrm{H}}^2$ is the halo (one-halo) contribution. These terms are given by:
\begin{align}
    \Delta_{\mathrm{Q}}^2(k) &= \Delta_{\mathrm{L}}^2(k)\,\frac{\left[1+\Delta_{\mathrm{L}}^2(k)\right]^{\beta_n}}{1+\alpha_n\,\Delta_{\mathrm{L}}^2(k)}e^{-f(y)}, \\
    \Delta_{\mathrm{H}}^2(k) &= \frac{\Delta_{\mathrm{H}}'^2(k)}{1 + \mu_n y^{-1} + \nu_n y^{-2}}, \\
    \Delta_{\mathrm{H}}'^2(k) &= \frac{a_n\, y^{3 f_1(\Omega_m)}}{1 + b_n\, y^{f_2(\Omega_m)} + \left[c_n\, f_3(\Omega_m)\, y\right]^{3-\gamma_n}},
\end{align}
where $\Delta_{\mathrm{L}}^2$ is the linear MPS and $f(y) \equiv y/4 + y^2/8$. The variable $y \equiv k/k_{\mathrm{NL}}$ depends on the nonlinear scale, $R_\text{NL} = 1/k_\text{NL}$, which is defined by $\sigma^2(R_{\mathrm{NL}})=1$ in Eq.~\eqref{Eq: sigma_R}. The functions $f_i(\Omega_m)$ are given by:
\begin{align}
    f_1(\Omega_m) &= \Omega_m^{-0.0307}, \\ 
    f_2(\Omega_m) &= \Omega_m^{-0.0585},\\
    f_3(\Omega_m) &= \Omega_m^{0.0743},
\end{align}
and the fitting coefficients read:
\begin{align}
    \log_{10} a_n &= 1.5222 + 2.8553 n_\eff + 2.3706 n_\eff^2 \nonumber \\
     &+ 0.9903 n_\eff^3 + 0.2250 n_\eff^4 - 0.6038 C, \\
    \log_{10} b_n &=
     -0.5642 + 0.5864 n_\eff + 0.5716 n_\eff^2 \nonumber \\
     &- 1.5474 C, \\
    \log_{10} c_n &=
     0.3698 + 2.0404 n_\eff + 0.8161 n_\eff^2 \nonumber \\
     &+ 0.5869 C, \\
    \gamma_n &= 0.1971 - 0.0843 n_\eff + 0.8460 C, \\
    \alpha_n &= \big| 6.0835 + 1.3373 n_\eff - 0.1959 n_\eff^2 \nonumber \\
     &- 5.5274 C \big|, \\
    \beta_n &= 2.0379 - 0.7354 n_\eff + 0.3157 n_\eff^2 \nonumber \\
     &+ 1.2490 n_\eff^3 + 0.3980 n_\eff^4 - 0.1682 C, \\
    \mu_n &= 0, \\
    \log_{10} \nu_n &= 5.2105 + 3.6902 n_\eff.
\end{align}
These coefficients depend on the effective spectral index, $n_\eff$, and the curvature, $C$, which are defined as:
\begin{align}
    n_\eff &= -3 -\left. \frac{d\ln\sigma^2(R)}{d\ln R}\right|_{R_{\rm NL}}, \\
    C &= -\left. \frac{d^2\ln\sigma^2(R)}{d(\ln R)^2}\right|_{R_{\rm NL}}.
\end{align}
In Fig.~\ref{Fig: Halofit}, we show a representative example of the NL MPS obtained with \texttt{halofit} when using either our emulator or \class~to supply the linear input. In this case, the fractional difference between the nonlinear spectra generated from the emulator and from \class~remains below $1\%$ down to $k \sim 8 \, h\,\Mpc^{-1}$. When tested against 100 NL spectra from \class, our emulator achieves a mean fractional error of $\sim 0.30\%$. A summary of the corresponding accuracy metrics for our interpretable model $P_\GA$, along with the variants that incorporate localized corrections, is presented in Table~\ref{Tab: Pnl metrics}.

\begin{center}
\begin{table}
\begin{centering}
\begin{tblr}{ c | c | c | c | c }
\hline[1.2pt]
\hline
Model & Mean($\%$) & Median($\%$) & $68\%$ & $95\%$ \\
\hline
\hline[1.2pt]
$P_\text{full} +$\texttt{halofit} & 0.30 & 0.28 & 0.32 & 0.45 \\
\hline
$P^{(\text{corr})}_\GA +$\texttt{halofit} & 0.34 & 0.33 & 0.36 & 0.49 \\
\hline
$P_\GA +$\texttt{halofit} & 0.42 & 0.41 & 0.45 & 0.54 \\
\hline[1.2pt]
\end{tblr}
\caption{Fractional-error statistics (mean, median, and central credible intervals) characterizing the accuracy of the nonlinear MPS predicted by \texttt{halofit} using our GA-derived linear spectra as input.}
\label{Tab: Pnl metrics}
\end{centering}
\end{table}
\end{center}

\section{Parametric Formula for the $P(k)$ of Modified Gravity}
\label{Sec: Parametric Formula for MG Theories}

\begin{figure*}[t!]
\centering    
\includegraphics[width=\columnwidth]{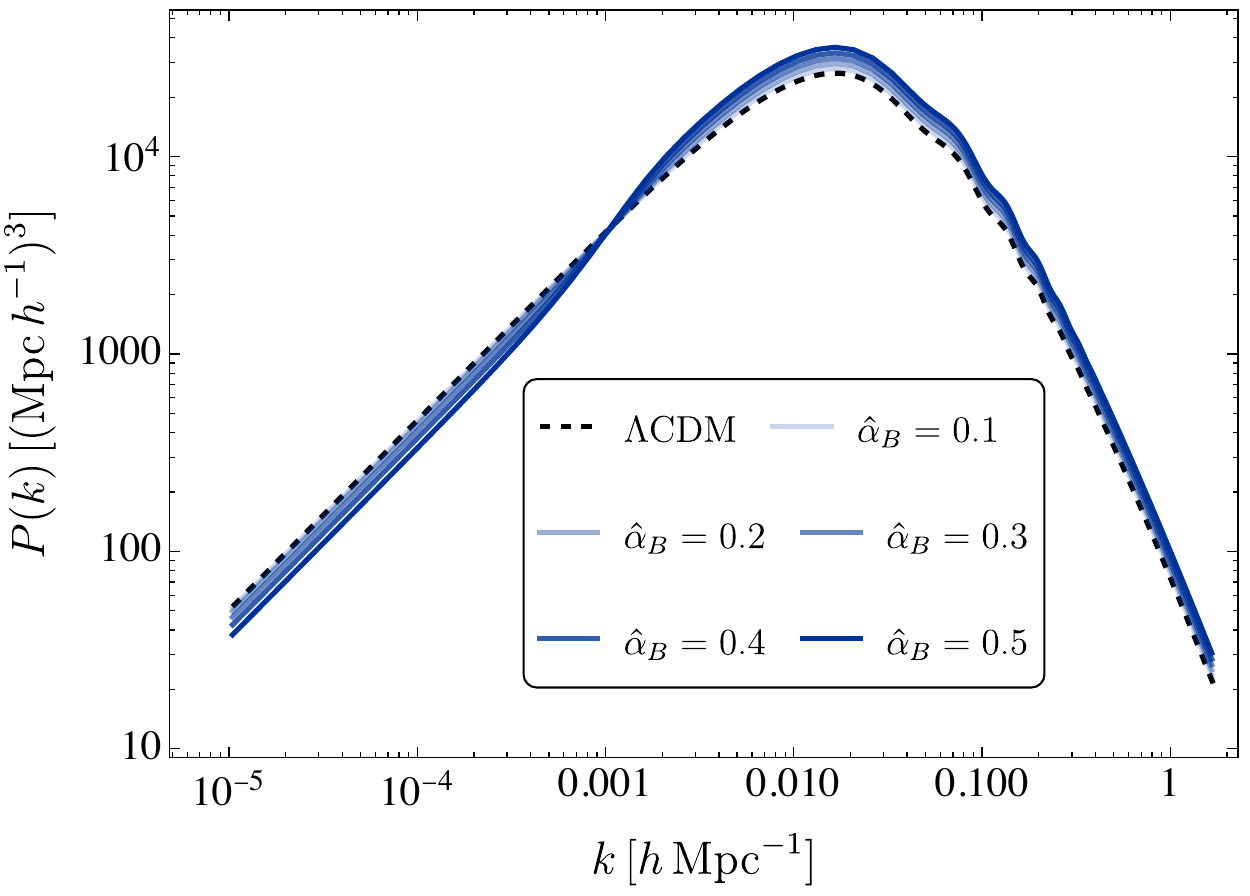} \hfill
\includegraphics[width=\columnwidth]{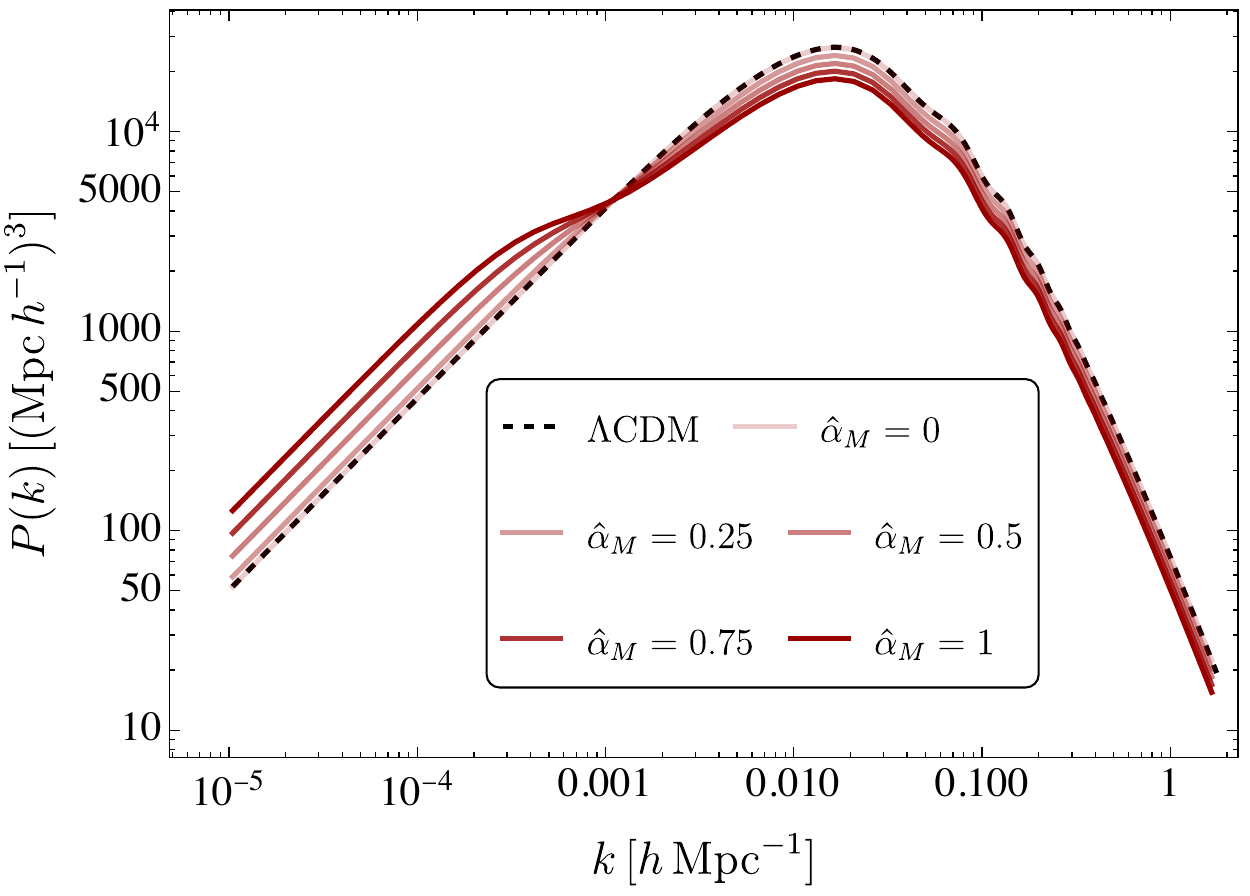}
\caption{MPS in the \texttt{propto\_scale} model of \texttt{hi\_class}, varying the braiding parameter $\hat{\alpha}_B$ (left) and the mass running parameter $\hat{\alpha}_M$ (right). The background evolution follows $\Lambda$CDM. \textbf{Left:} MG suppresses power at large scales and enhances it at smaller scales. \textbf{Right}: the opposite trend is observed. A transition is visible near the scale of matter–radiation equality.}
\label{Fig: MG P(k)}
\end{figure*}

In Ref.~\cite{Orjuela-Quintana:2023uqb}, we introduced a parametric expression designed to encapsulate various effects arising from modifications to GR. In this work, we revisit and update that formulation by incorporating the new expression we have derived for the smooth MTS.

We begin by identifying the main ways in which MG theories can affect the MPS relative to the standard $\Lambda$CDM scenario:
\begin{enumerate}
    \item MG may introduce a scale-dependent growth factor.
    \item MG may lead to effective dark energy clustering during structure formation~\cite{Sapone:2009mb, Sapone:2012nh, Sapone:2013wda}.
    \item Deviations at the largest scales may emerge due to the integrated Sachs–Wolfe (ISW) effect.
\end{enumerate}

To illustrate these effects, we compute the MPS in two representative MG scenarios, varying key parameters to visualize their physical impact. The resulting spectra are shown in Fig.~\ref{Fig: MG P(k)}, obtained with the \texttt{hi\_class} extension of \class~\cite{Zumalacarregui:2016pph, Bellini:2019syt}. We consider the built-in \texttt{propto\_scale} model, where each effective field theory (EFT) function that characterizes the dynamics of Horndeski theories is proportional to the scale factor. For an overview of these EFT functions, see Refs.~\cite{Bellini:2014fua, Bellini:2015xja}.

In the left and right panels of Fig.~\ref{Fig: MG P(k)}, we vary the proportionality constants of the braiding parameter $\hat{\alpha}_B$ and the mass running parameter $\hat{\alpha}_M$, respectively, while keeping the background evolution fixed to that of $\Lambda$CDM. We observe that MG effects introduce scale-dependent modifications to $P(k)$: the form of the spectrum remains nearly unchanged at large scales (following $P(k) \propto k^{n_s}$) except for a shift in its amplitude, while deviations—either enhancements or suppressions—appear at smaller scales, with a transition typically occurring before $k_\eq$. The $\Lambda$CDM spectrum is shown for reference in all panels.

\subsection*{Parametric Model}

We now introduce a parametrized formula that captures the MG-induced effects on the MPS. The model is designed to satisfy two main goals:
\begin{enumerate}
    \item Modify the normalization of $P(k)$ across scales, encoding large-scale ISW effects and small-scale scale-dependent growth.
    \item Introduce suppression or enhancement of power at intermediate and small scales due to dark energy clustering or fifth forces.
\end{enumerate}

To achieve this, we define two separate spectral components:
\begin{itemize}
    \item At large scales, where the MPS retains a primordial shape, we define:
    \begin{equation}
    \label{Eq: Plarge}
        P_\text{large} (k) \equiv A_0^{(\MG)} (1 + s_\MG)\, k^{n_s},
    \end{equation}
    where $s_\MG$ accounts for deviations due to the ISW effect or background evolution, and:
    \begin{equation}
        A_0^{(\MG)} \equiv 2\pi^2 \left( \frac{A_s}{10^{9}} \right)\left(\frac{h}{k_\text{p}}\right)^{n_s - 1}\left( \frac{2 \, c^2 h^2}{5 \, \omega_m} \right)^2,
    \end{equation}
    is the normalization constant from Eq.~\eqref{Eq: General MPS}, valid in the subhorizon regime. Here, $c$ is the speed of light in units of $\text{km}\,\text{s}^{-1}$.
    
    \item At smaller scales, we define a second component that includes a suppression/enhancement term:
    \begin{align}
    \label{Eq: Psmall}
        P_\text{small} (k) &= A_0^{(\MG)} \, k^{n_s}\, T_\nw^2(k) \nonumber \\
        &\quad \times \left(1 + 0.711\, \alpha_\MG^3 + 1.88\, \alpha_\MG^4 + 0.939\, \alpha_\MG^5 \right) \nonumber \\
        &\quad \times \left(1 + \frac{\gamma_\MG}{1 + (k_\MG/k)^{\sigma_\MG}} \right).
    \end{align}
    The first parenthesis mimics a rescaled amplitude, while the second introduces a scale-dependent modification motivated by MG theories. The parameter $\gamma_\MG$ controls the strength of the deviation, $k_\MG$ sets its characteristic scale, and $\sigma_\MG$ determines its sharpness.
\end{itemize}

This functional form is physically motivated by modifications to the Poisson equation in MG scenarios. For example, in $f(R)$ gravity under the quasi-static approximation~\cite{Cardona:2022lcz, Orjuela-Quintana:2023zjm}:
\begin{align}
    k^2 \Phi &= - 4\pi G [1 + \mu(k)]\, \bar{\rho}_m\, \delta_m, \\
    \mu(k) &\equiv \frac{1/3}{1 + (a\, m(a)/k)^2},
\end{align}
where $\mu(a, k)$ is the slip parameter, and $m(a)$ is the scalaron mass. In this case, we recover our parametric correction at $a = 1$ with:
$$
\gamma_\MG = \frac{1}{3}, \quad k_\MG = m_0, \quad \sigma_\MG = 2.
$$

\begin{center}
\begin{table*}[t]
\begin{centering}
\begin{tblr}{X[c] | X[c] | X[c] | X[1.5, c] | X[c] | X[c] | X[c] | X[c] | X[1.5, c] | X[c]}
\hline[1.2pt]
\hline
$E_{11}$ & $E_{22}$ & $\alpha_\MG$ & $k_\text{T} \times 10^{-3}$ & $\beta_\text{T}$ & $s_\MG$ & $\gamma_\MG$ & $k_\MG$ & $\sigma_\MG \times 10^{-3}$ & $\mape$ \\
\hline
\hline[1.2pt]
$1.0$ & $1.0$ & $-0.779$ & $1.039$ & $0.472$ & $-0.894$ & $-0.673$ & $1.028$ & $-4.933$ & $1.44\%$ \\
\hline
$-1.0$ & $-1.0$ & $-0.757$ & $0.114$ & $0.477$ & $23.90$ & $-1.031$ & $1.488$ & $1.098$ & $1.45\%$ \\
\hline
$0.5$ & $0.5$ & $-0.741$ & $0.766$ & $0.494$ & $-0.771$ & $-0.753$ & $1.136$ & $-1.617$ & $1.41\%$ \\
\hline
$-0.5$ & $-0.5$ & $-0.776$ & $0.281$ & $0.513$ & $1.447$ & $-0.963$ & $1.437$ & $2.132$ & $1.42\%$ \\
\hline[1.2pt]
\end{tblr}
\caption{Best-fit MG parameters used to construct the smooth spectra $P_\nw(k)$ for the \texttt{plk\_late} model, with parameters $E_{11}$ adn $E_{22}$, as obtained from \texttt{mg\_class}. The mean absolute percentage error is around $\sim1.5\%$.}
\label{Tab: MG Params plk late}
\end{centering}
\end{table*}
\end{center}

To combine both regimes, we introduce a smooth transition function:
\begin{equation}
\label{Eq: transition function}
    \sigma_\text{T}(k) \equiv \left[ 1 + e^{-(\ln k - \ln k_\text{T})/\beta_\text{T}} \right]^{-1},
\end{equation}
which interpolates between \( P_\text{large} \) at \( k \ll k_\text{T} \) and \( P_\text{small} \) at \( k \gg k_\text{T} \), with $\beta_\text{T}$ controlling the smoothness.

The final parametric formula for the matter power spectrum in MG theories is:
\begin{equation}
\label{Eq: P(k) of MG}
    P_\text{MG}(k| \boldsymbol{\theta}_\MG) = 
    (1 - \sigma_\text{T}) \, P_\text{large}(k) + 
    \sigma_\text{T}\, P_\text{small}(k),
\end{equation}
where the full parameter set is given by:
$$
\boldsymbol{\theta}_\MG = \{ \alpha_\MG, \beta_\text{T}, k_\text{T}, s_\MG, \gamma_\MG, k_\MG, \sigma_\MG \}.
$$

This model introduces seven MG-specific parameters:
\begin{itemize}
    \item $s_\MG$: Large-scale normalization shift (e.g., ISW effect),
    \item $\alpha_\MG$: Small-scale amplitude modification,
    \item $k_\text{T}, \beta_\text{T}$: Transition scale and width,
    \item $\gamma_\MG$: Strength of suppression/enhancement at small scales,
    \item $k_\MG$: Characteristic MG scale,
    \item $\sigma_\MG$: Width of MG modification.
\end{itemize}
Each parameter is physically motivated and interpretable. 

In order to illustrate the flexibility of our parametric deformation of the baseline emulator, we consider a representative MG model that exhibits substantial deviations from the standard $\Lambda$CDM spectrum across all scales. Specifically, we focus on the \texttt{plk\_late} model, in which the gravitational slip parameters are expressed as functions of $\Omega_\text{DE}$ (see Ref.~\cite{Sakr:2021ylx}) and governed by two coefficients, $E_{11}$ and $E_{22}$. This parametrization has been used in observational analyses, including those of the Planck~\cite{Planck:2015bue} and DES~\cite{DES:2018ufa} collaborations.

We compute the linear MPS using \texttt{mg\_class}, where this MG model is already implemented. For each set of MG parameters, we then perform a simple numerical minimization of the \mape, keeping the baseline cosmology fixed to the default \class~values:
\begin{align}
\label{Eq: class default}
h &= 0.6781, &
\omega_b &= 0.02238, &
\omega_c &= 0.1201, \nonumber \\
n_s &= 0.9660, &
A_s &= 2.1005 \times 10^{-9}.
\end{align}
We vary only the parameters $E_{11}$ and $E_{22}$. The minimization yields the values reported in Table~\ref{Tab: MG Params plk late}, resulting in an average fractional error of approximately $1.5\%$.

\begin{figure}[t!]
\centering    
\includegraphics[width=\columnwidth]{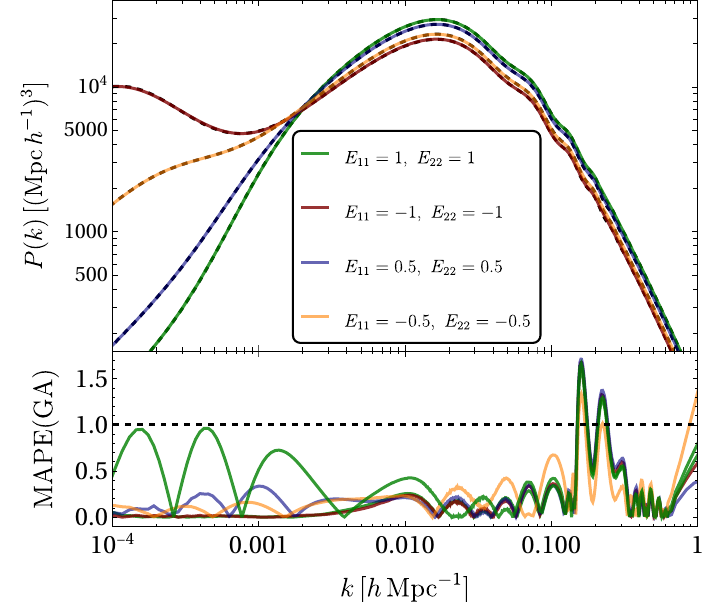}
\caption{Comparison between the linear MPS for the \texttt{plk\_late} models computed using \texttt{mg\_class} and our parametric formula. \textbf{Top:} MPS obtained with \texttt{mgclass} (black dashed curves) alongside the corresponding predictions from our parametric formula (solid curves) for the best-fit values of $(E_{11}, E_{22})$ listed in Table~\ref{Tab: MG Params plk late}. \textbf{Bottom:} \mape~as a function of $k$. Our parametric formula accurately reproduces the spectra from this MG model, achieving a fractional error below $1\%$ across most scales once the wiggly MTS contribution is included.}
\label{Fig: plk late}
\end{figure}

As shown in Fig.~\ref{Fig: plk late}, our parametric formula successfully reproduces the significant large-scale departures from $\Lambda$CDM characteristic of the \texttt{plk\_late} spectra. Remarkably, once the wiggly MTS is incorporated into the MG prescription, the fractional error drops below $1\%$ over nearly the entire range of scales.

In the following section, we investigate how this parametric formulation can be employed to analyze the impact of MG effects on the position of the BAO peak.

\section{Contact with Observations: The BAO Scale}
\label{Sec: BAO Scale}

\begin{center}
\begin{table*}[t]
\begin{centering}
\begin{tblr}{X[c] | X[1.5, c] | X[c] | X[1.5, c] | X[c] | X[c] | X[c] | X[c] | X[c] | X[c]}
\hline[1.2pt]
\hline
Model & $f_{R_0}$ & $\alpha_\MG$ & $k_\text{T} \times 10^{-3}$ & $\beta_\text{T}$ & $s_\MG$ & $\gamma_\MG$ & $k_\MG$ & $\sigma_\MG$ & $\mape$ \\
\hline
\hline[1.2pt]
$f_{R_\text{p}}$ & $5 \times 10^{-7}$ & $-0.367$ & $0.676$ & $0.441$ & $-0.378$ & $-0.376$ & $1.852$ & $-1.386$ & $1.56\%$ \\
\hline
$f_{R_\text{pp}}$ & $5 \times 10^{-6}$ & $-0.368$ & $0.653$ & $0.439$ & $-0.378$ & $-0.376$ & $0.642$ & $-1.263$ & $1.52\%$ \\
\hline
$f_{R_\text{ppp}}$ & $5 \times 10^{-5}$ & $-0.084$ & $1.494$ & $0.524$ & $-0.378$ & $-0.401$ & $0.242$ & $-0.986$ & $1.47\%$ \\
\hline
$f_{R_\text{pppp}}$ & $5 \times 10^{-4}$ & $0.661$ & $2.296$ & $0.511$ & $-0.378$ & $-0.675$ & $0.659$ & $-0.621$ & $1.85\%$ \\
\hline[1.2pt]
\end{tblr}
\caption{Best-fit MG parameters used to construct the smooth spectra $P_\nw(k)$ for different values of $f_{R_0}$ in the Hu-Sawicki model as obtained from \texttt{mg\_class}. The average fractional error is around $\sim1.5\%$, sufficient for BAO analyses having into account that SG filter on $\Lambda$CDM achieves $\sim1\%$.}
\label{Tab: MG Params}
\end{centering}
\end{table*}
\end{center}

Although the MPS is a cornerstone of modern cosmology, it functions primarily as a summary statistic. In practice, the main observable in LSS surveys is the two-point correlation function (2PCF), defined as:
\begin{equation}
    \xi(\mathbf{r}) \equiv \langle \delta(\mathbf{x})\delta(\mathbf{x} + \mathbf{r}) \rangle,
\end{equation}
where $\delta(\mathbf{x})$ denotes the matter overdensity at position $\mathbf{x}$. The 2PCF quantifies the clustering strength of matter as a function of spatial separation $\mathbf{r}$, providing complementary insights into structure formation~\cite{Dodelson:2020bqr}. The MPS and 2PCF are Fourier counterparts, with the following relation:
\begin{equation}
    \xi(\mathbf{r}) = \frac{1}{(2\pi)^3} \int \text{d}^3 k \, P(\mathbf{k}) \, e^{- i \mathbf{k} \cdot \mathbf{r}}.
\end{equation}

In this section, we compute the 2PCF using our analytic expressions for $P(k)$ and compare the resulting BAO templates with those obtained via conventional approaches, with particular attention to the amplitude and position of the BAO peak.

In LSS surveys (e.g., SDSS, BOSS), the observed MPS is modeled as~\cite{BOSS:2016apd, eBOSS:2020lta}:
\begin{align}
P_\text{obs}(k| \mu) &= C(k, \mu) \, P_\nw(k) \\ &\times \left[1 + \left\{O_\text{lin}(k) - 1\right\} e^{-k^2 \Sigma_\text{nl}^2/2} \right],
\end{align}
where $P_\nw(k)$ is the de-wiggled spectrum, and $O_\text{lin} \equiv P_\text{lin}/P_\nw$ captures the oscillatory BAO signal. The term $C(k, \mu)$ encodes redshift-space distortions and Finger-of-God effects:
\begin{equation}
C(k, \mu) = \frac{\left[\sigma_8 b + f\sigma_8\{1 - S(k)\}\mu^2\right]^2}{1 + k^2 \mu^2 \Sigma_s^2 / 2},
\end{equation}
where $b$ is the linear bias, $f\sigma_8$ the growth rate, $S(k) = e^{-k^2 \Sigma_r^2/2}$ is the reconstruction smoothing function with $\Sigma_r = 15 \, \Mpc\,h^{-1}$, and $\Sigma_s$ characterizes small-scale random motions. The nonlinear damping is modeled by:
\begin{align}
\Sigma^2_\text{nl}(\mu) &= \frac{2}{3} \Sigma_{xy}^2 + \frac{1}{3} \Sigma_z^2, \\
\Sigma_{xy} &= \sigma_8 \int \frac{\dd k}{2\pi^2} k \, P_\text{lin}(k), \\
\Sigma_z &= (1 + f\sigma_8) \Sigma_{xy}.
\end{align}
Finally, the Legendre multipoles of the power spectrum are computed as:
\begin{equation}
P_\ell(k) = \frac{2\ell + 1}{2} \int_{-1}^{1} \text{d}\mu \, P_\text{obs}(k; \mu) L_\ell(\mu),
\end{equation}
and the corresponding multipoles of the 2PCF follow via spherical Bessel transforms:
\begin{equation}
\xi_\ell(r) = \frac{i^\ell}{2\pi^2} \int \dd k\, k^2 P_\ell(k) j_\ell(kr).
\end{equation}
where $L_\ell$ are the Legendre polynomials and $j_\ell$ are the spherical Bessel functions.

\begin{figure*}[t!]
\centering    
\includegraphics[width=\columnwidth]{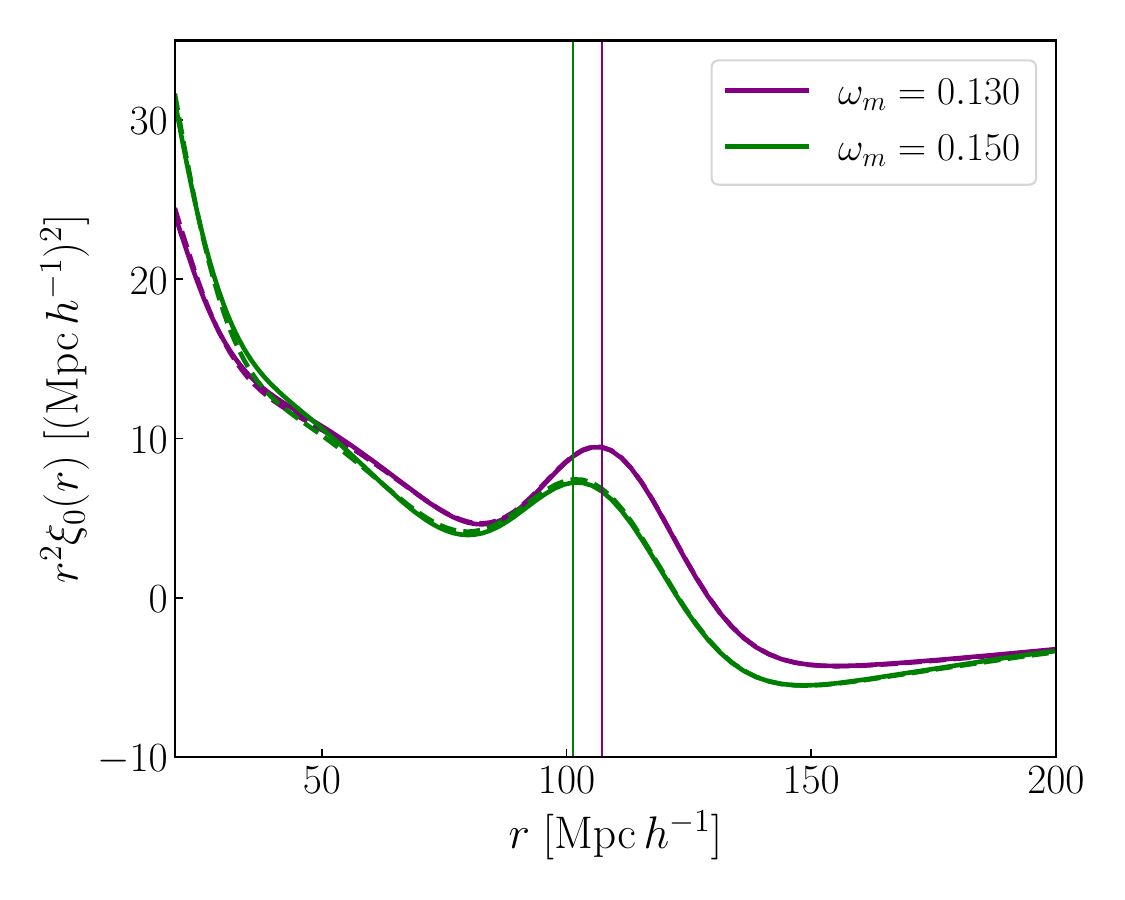} \hfill
\includegraphics[width=\columnwidth]{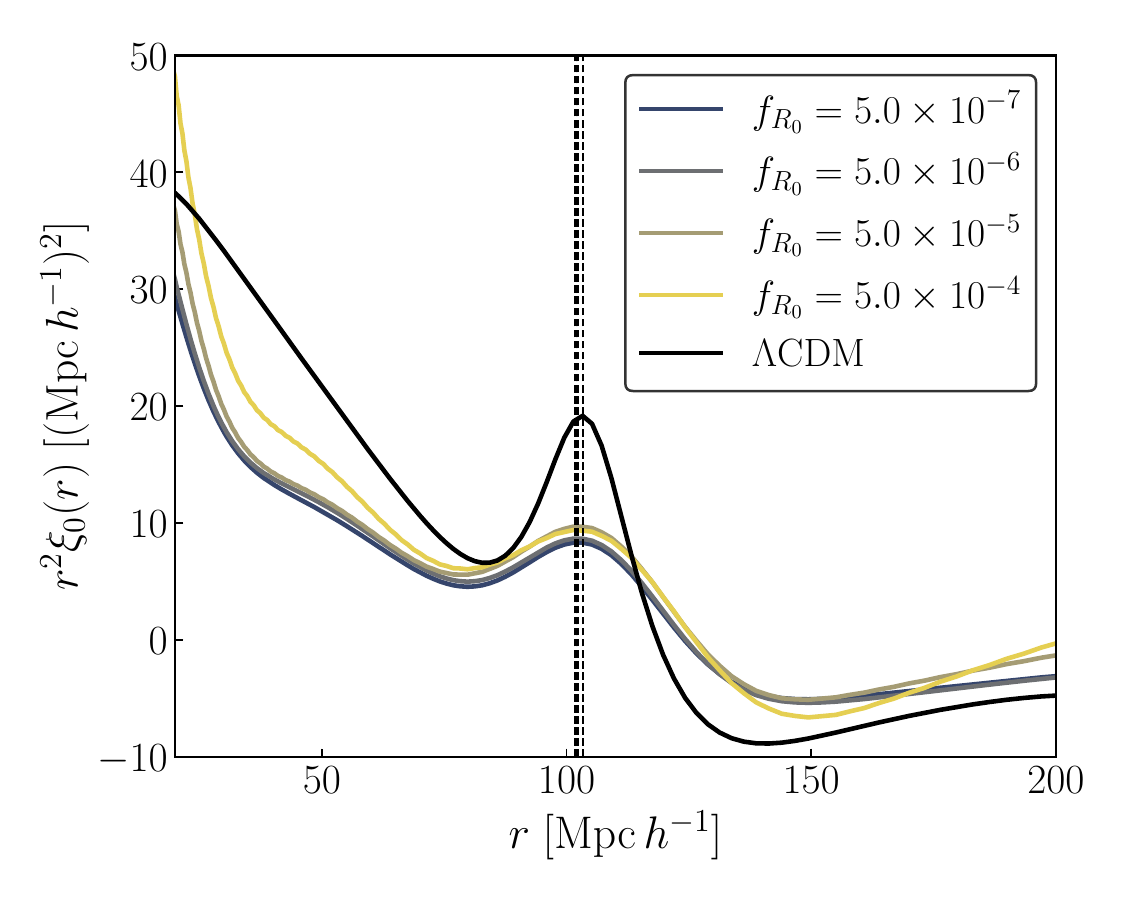}
\caption{\textbf{Left:} Comparison of the monopole $r^2\xi_0$ computed using \class~+ EH (dashed) and our only GA-based formula (solid) for two values of $\omega_m$. Vertical lines mark the BAO peak position. \textbf{Right:} BAO template computed for four values of $f_{R_0}$ in the Hu-Sawicki $f(R)$ model using our MG-parametrized smoothing. The BAO peak position and amplitude shows slight sensitivity to $f_{R_0}$ at the linear level.}
\label{Fig: BAO 2PCF}
\end{figure*}

\begin{table*}
\begin{centering}
\begin{tblr}{ c | c | c | c | c | c }
\hline[1.2pt]
\hline
$f_{R_0}$ & $0$ & $5\times10^{-7}$ & $5\times10^{-6}$ & $5\times10^{-5}$ & $5\times10^{-4}$ \\
\hline
\hline[1.2pt]
Peak position $[\Mpc\,h^{-1}]$ & 103.28 & 102.33 & 102.22 & 102.09 & 101.72 \\
\hline
Peak shift $[\Mpc\,h^{-1}]$ & 0.0 & 0.95 & 1.06 & 1.19 & 1.56 \\
\hline[1.2pt]
\end{tblr}
\caption{Position of the BAO peak for the $\Lambda$CDM model and four realizations of the parameter $f_{R_0}$. A stronger gravity compared with $\Lambda$CDM ($f_{R_0} = 0$) shifts the BAO peak towards smaller scales.}
\label{Tab: BAO shift}
\end{centering}
\end{table*}

In conventional analyses, $P_\text{lin}$ is obtained from Boltzmann solvers, and $P_\nw$ is estimated using the EH fitting formula or smoothing techniques like SG filtering. For comparison, we will use our GA-based expressions for $P_{\GA, \nw}$ and $P_\MG$. This is particularly relevant in light of recent results~\cite{Chen:2024tfp}, which show that the de-wiggling procedure can significantly affect the extracted BAO signal. However, the corresponding the BAO scale seems to be rather insensitive to this methodology choice.

\subsection{$\Lambda$CDM Case}

The left panel of Fig.~\ref{Fig: BAO 2PCF} shows the monopole moment $r^2 \xi_0$ for $\omega_m = 0.13$, and $\omega_m =0.15$, while other cosmological parameters are fixed to their \class~default values in Eq.~\eqref{Eq: class default}, computed using \class~with the EH de-wiggling prescription (dashed lines), and using our GA-derived MPS (solid lines). We set $b = 1$ at $z=0$, and extract $\sigma_8$ and $f\sigma_8$ from \class. As expected, increasing $\omega_m$ shifts the BAO peak to smaller scales. Although both de-wiggling functions yield different BAO signals, their location and amplitude of the BAO peak are rather indistinguishable, as indicated by the vertical lines.  

\subsection{$f(R)$ Case}

To explore the impact of gravity modifications on the BAO scale, we analyze the Hu-Sawicki $f(R)$ model under the assumption of a $\Lambda$CDM-like background~\cite{Hu:2007pj, Orjuela-Quintana:2023zjm}. This model is governed by a single parameter $f_{R_0}$, which quantifies deviations from GR. Following the \texttt{Quijote-MG} simulation setup in Ref.~\cite{Valogiannis:2024rvt},\footnote{\href{https://quijote-simulations.readthedocs.io/en/latest/mg.html}{https://quijote-simulations.readthedocs.io/en/latest/mg.html}} we fix the background cosmology to the \texttt{Quijote fiducial} model and vary:
\begin{align}
&\{ f_{R_\text{p}}, f_{R_\text{pp}}, f_{R_\text{ppp}}, f_{R_\text{pppp}} \} = \\
&\{5\times10^{-7}, 5\times10^{-6}, 5\times10^{-5}, 5\times10^{-4} \}. \nonumber
\end{align}

We compute the MPS using \texttt{mg\_class}~\cite{Sakr:2021ylx} and apply our MG-parametrized GA expression to extract $P_\nw(k)$. Table~\ref{Tab: MG Params} summarizes the best-fit parameters obtained from this smoothing procedure.

The right panel of Fig.~\ref{Fig: BAO 2PCF} shows the resulting BAO templates. As shown in Table~\ref{Tab: BAO shift}, we find that the BAO peak position is largely unaffected by the value of $f_{R_0}$, indicating that the linear BAO scale is relatively robust to moderate deviations from GR, showing a shift of $\sim 1.56~[\Mpc\,h^{-1}]$ from the $\Lambda$CDM case for $f_{R_0} = 5 \times 10^{-4}$ which is a rather large value for this parameter. However, it is important to note that this analysis is limited to linear theory. Nonlinear and mildly nonlinear effects during structure formation can impact the BAO scale more significantly, especially in MG scenarios. A complete treatment of such effects, though beyond the scope of this work, is crucial for next-generation surveys and remains an open area of research.

\section{Parameter Identifiability and Degeneracies}
\label{Sec: Fisher Analysis}

\begin{figure*}[t]
    \centering
    {\includegraphics[width=\columnwidth]{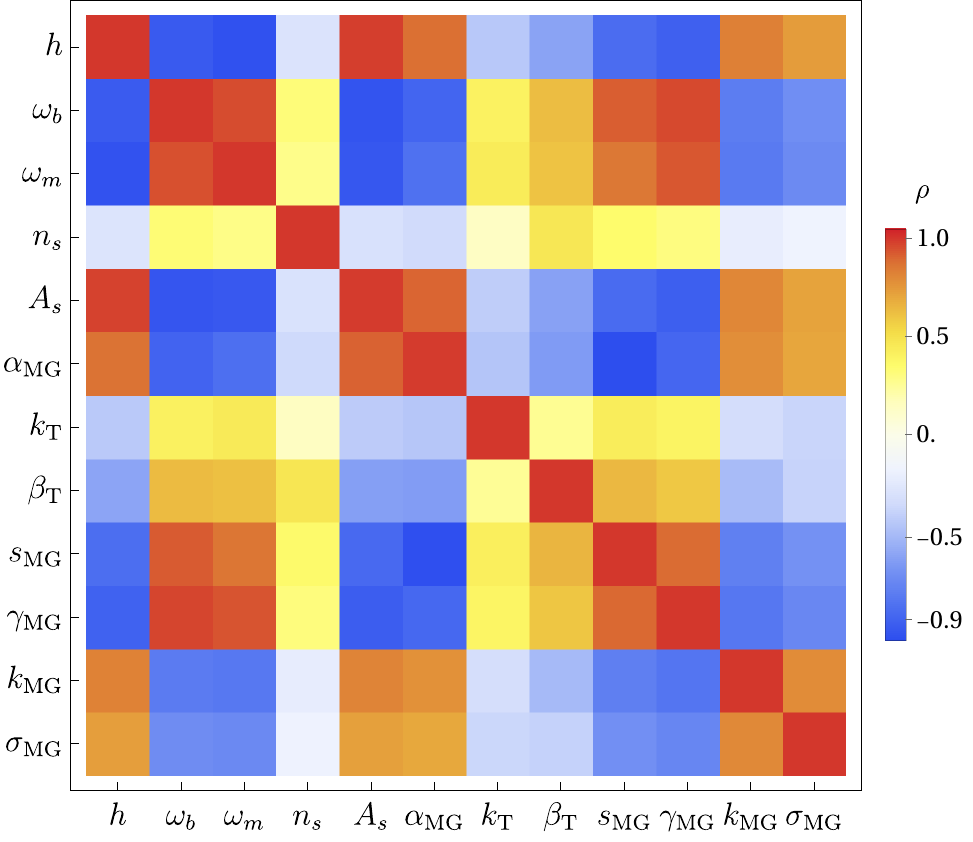} \hfill
    \includegraphics[width=\columnwidth]{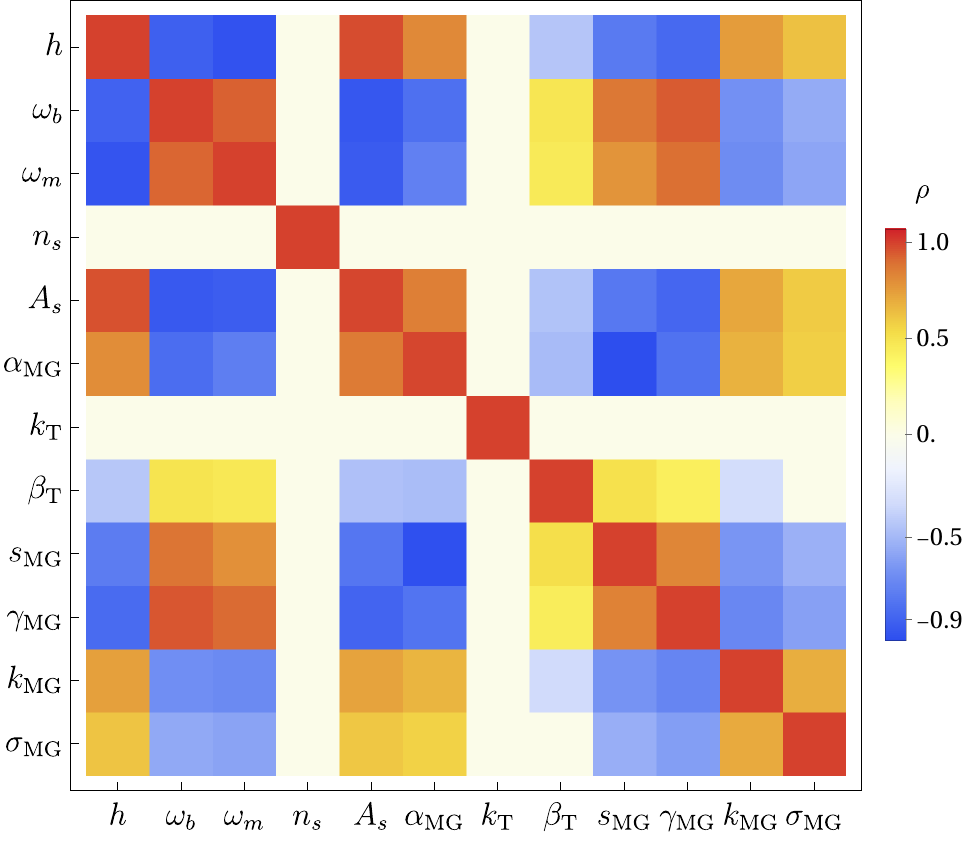}}
    \caption{\textbf{Left:} Correlation matrix for the full set of cosmological and MG parameters evaluated at the best-fit $f_{R_\text{p}}$ solution. \textbf{Right:} The same matrix with a mask filtering $|\rho| < 0.6$, highlighting only the strongest correlations. While amplitude-like parameters such as $A_s$ and $\alpha_\MG$ show expected degeneracies, the transition scale $k_{\text{T}}$ and spectral index $n_s$ remain largely decoupled.}
    \label{Fig: Full Correlation}
\end{figure*}

A central concern in any extensive parameterization of modified gravity is the potential for degeneracy, where different combinations of parameters may mimic the same phenomenological features. To assess the identifiability of the model parameters and quantify potential degeneracies, we perform a Fisher Matrix analysis around the best-fit solution found for the $f_{R_\text{p}}$ model. However, our approach can be extended to other sets of parameters.

The Fisher information matrix, $F_{ij}$, is computed as:
\begin{equation}
    F_{ij} = \sum_{n} \frac{1}{\sigma_{P}^2(k_n)} \frac{\partial P_\MG(k_n)}{\partial \theta_i} \frac{\partial P_\MG(k_n)}{\partial \theta_j} w_n,
\end{equation}
where $\theta$ represents the set of 12 parameters (5 cosmological + 7 MG), and the derivatives were evaluated at the best-fit values provided in Table~\ref{Tab: MG Params}. To focus on the structural degeneracies of the functional form, we adopt a variance-limited assumption where the error is proportional to the signal, $\sigma_P(k) \propto P_{\text{MG}}(k)$. The sum runs over the broadband range of wavenumbers used in our analysis, with weights $w_n$ accounting for the logarithmic spacing. The covariance matrix, $C_{ij}$, is then derived from the inverse Fisher matrix and, subsequently, the correlation matrix is defined in terms of the covariance matrix as:
\begin{equation}
    \mathcal{R}_{ij} = \frac{C_{ij}}{\sqrt{C_{ii} C_{jj}}}.
\end{equation}

\subsection{Global Degeneracies}

Left panel of Fig.~\ref{Fig: Full Correlation} displays the correlation matrix for the full parameter set, including both standard cosmological parameters ($h, \omega_b, \omega_m, n_s, A_s$) and the 7 MG parameters. The right panel shows the same plot considering a threshold in the density $|\rho| > 0.6$, and thus the strongest correlated parameters are shown.

As anticipated for a flexible broadband parameterization, we observe strong correlations between parameters governing the global amplitude. Specifically, the primordial amplitude $A_s$ is degenerate with the MG parameters $\alpha_\MG$ and $s_\MG$, which also regulate the overall power normalization. Similarly, the MG shape parameters show partial degeneracies with the Hubble parameter $h$.

However, a crucial result from this analysis is the decoupling of the spectral index and the transition scale. As shown in the right panel of Fig.~\ref{Fig: Full Correlation}, $n_s$ and $k_\text{T}$ exhibit low correlation with the rest of the parameter space (regions of ivory). This indicates that the \textit{tilt} of the primordial spectrum and the physical \textit{location} of the MG transition are uniquely identifiable features in our parameterization, robust against amplitude degeneracies.

\subsection{Internal Structure of the MG Parametrization}

\begin{figure*}[t]
    \centering
    {\includegraphics[width=\columnwidth]{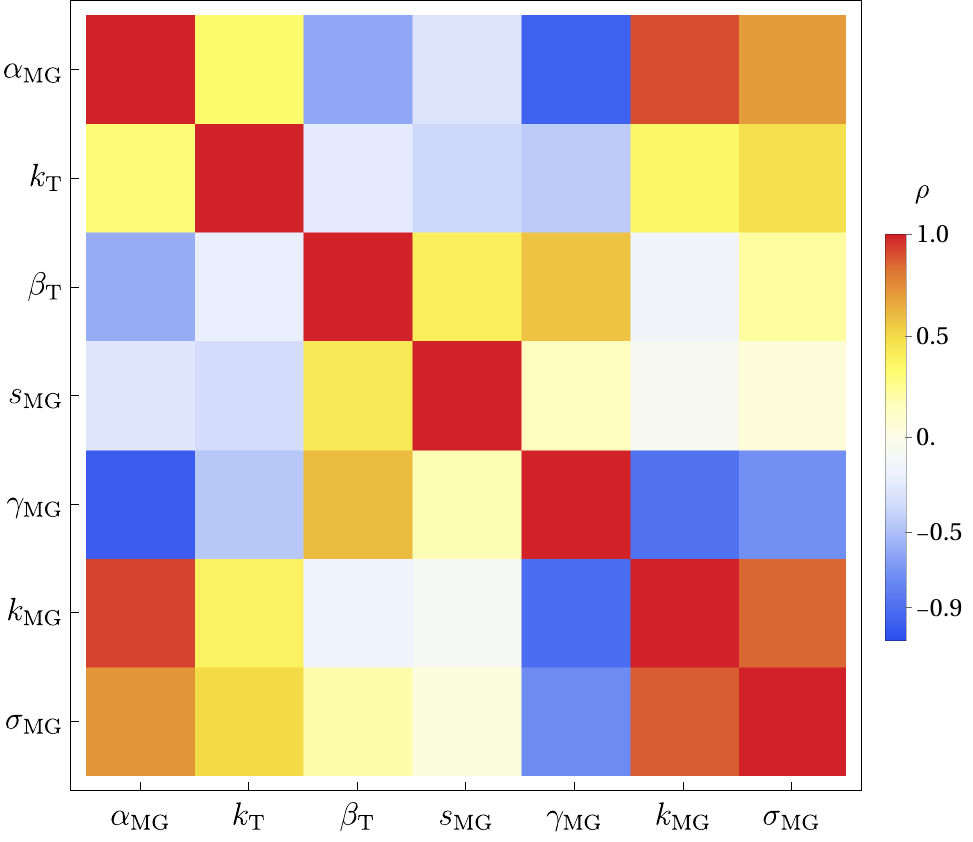} \hfill
    \includegraphics[width=\columnwidth]{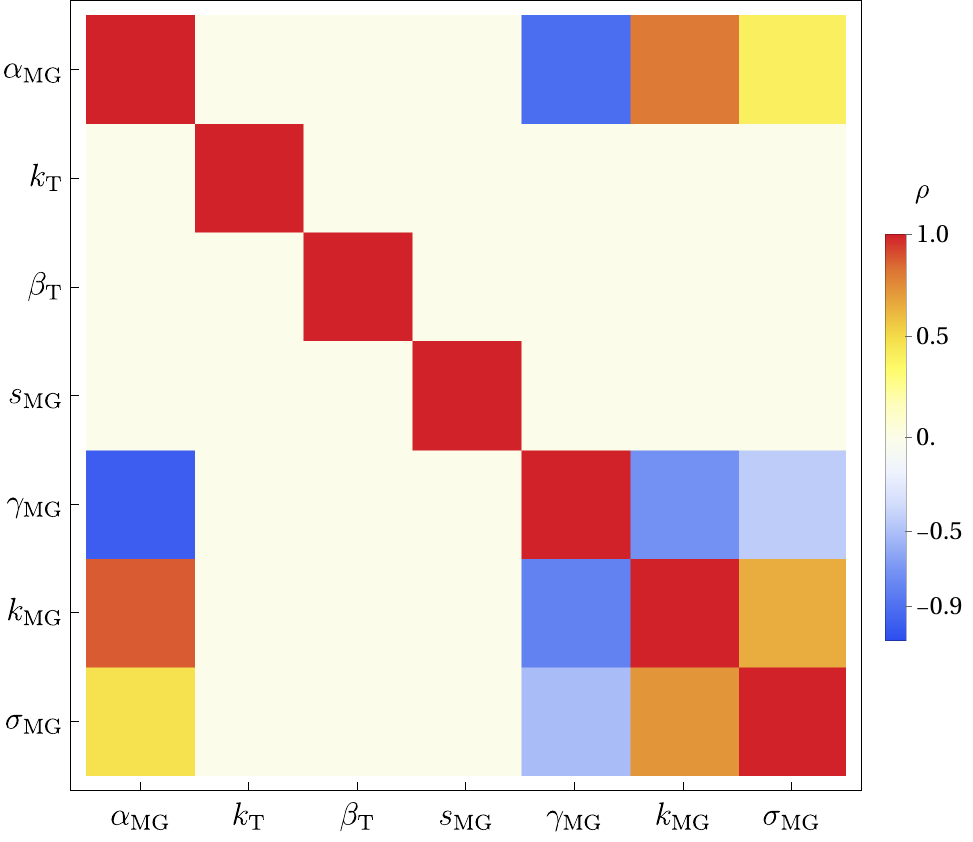}}
    \caption{\textbf{Left:} Internal correlation matrix for the MG parameters with a fixed cosmological background. \textbf{Right:} Masked matrix ($|\rho| < 0.6$) highlighting strong degeneracies. The analysis reveals three distinct sectors: a decoupled transition block ($k_{\text{T}}, \beta_{\text{T}}$), an independent linear term ($s_\MG$), and a degenerate block of small-scale shape parameters.}
    \label{Fig: Masked Correlation}
\end{figure*}

To further isolate the degeneracies intrinsic to the MG sector and guide the physical interpretation of the parameters, we performed a second Fisher analysis fixing the standard cosmological parameters to their fiducial values. This scenario emulates a practical application where the background cosmology is tightly constrained by external probes.

The resulting correlation matrix for the 7 MG parameters is shown in Fig.~\ref{Fig: Masked Correlation}. We applied a mask to highlight only significant correlations ($|\rho| > 0.6$). This analysis reveals a modular structure in the parametrization:
\begin{itemize}
    \item \textbf{Decoupled Transition Sector:} The transition parameters, $k_{\text{T}}$ and $\beta_{\text{T}}$, form an isolated cluster. They are correlated with each other but effectively decoupled from the shape parameters. This implies that the scale at which gravity departs from GR is a robust observable, unaffected by the modeling of the non-linear tail.
    
    \item \textbf{Unique Linear Modification:} The parameter $s_\MG$, which governs the linear large-scale modification, is uncorrelated with the non-linear shape parameters. This confirms that the model correctly separates linear growth suppression/enhancement from non-linear screening effects.
    
    \item \textbf{Shape Degeneracy Block:} The parameters governing the detailed small-scale shape ($\alpha_{\text{MG}}, \gamma_{\text{MG}}, k_{\text{MG}}, \sigma_{\text{MG}}$) exhibit strong internal correlations. This reflects a functional degeneracy: variations in the polynomial amplitude $\alpha_\MG$ can be partially compensated by adjustments in the rational function parameters.
\end{itemize}

Consequently, while the \textit{collective} small-scales shape is well-described, individual constraints on the shape parameters should be interpreted jointly. For robust physical inference, we recommend treating the shape block as nuisance parameters that marginalize over the uncertainty of the non-linear spectrum, while focusing on $k_{\text{T}}$, $\beta_\MG$, and $s_\MG$ as the primary physical indicators.

\section{Conclusions and Outlooks} 
\label{Sec: Conclusions}

In this work, which builds upon previous efforts~\cite{Orjuela-Quintana:2023uqb, Orjuela-Quintana:2022nnq}, we have developed an interpretable emulator for the linear matter power spectrum based on a ``\textit{physics-informed}'' approach to symbolic regression using genetic algorithms. The resulting framework combines analytic transparency with practical flexibility, capturing the main physical dependencies of the MPS while allowing for minimal empirical corrections near key transition scales.

We demonstrated that the smooth (non-wiggly) component of the transfer function can be described by a symbolic formula that closely mirrors the structure of the traditional BBKS approximation, but achieves a higher accuracy and lower complexity than this one, even improving over the widely utilized zero-baryon case of the Eisenstein-Hu formulation. This smooth component achieves performance comparable to numerical de-wiggling techniques such as the Savitzky-Golay filter, with the added benefit of being expressed in closed form.

Building upon this baseline, we constructed an analytical representation for the BAO wiggles that incorporates key physical effects—such as acoustic oscillations, Silk damping, and amplitude suppression—resulting in sub-percent level accuracy across the full relevant range of scales. Further corrections, including localized and skew-normal terms, were introduced to improve the agreement with numerical results around physically meaningful scales such as $k_\eq$ and $k_\Silk$.

Our physically interpretable emulator attains a final mean absolute percentage error of $\mape(\GA)=0.42\%$, while the version including localized corrections reaches $\mape(\GA) = 0.29\%$. Both outperform traditional fitting formulas such as Eisenstein--Hu and approach the precision of recent symbolic regression-based models~\cite{Sui:2024wob}, while maintaining a significantly lower symbolic complexity. The fractional error of our formula remains below 1\% across the scale range $k \in [10^{-5}, 1.5]~h\,\Mpc^{-1}$, thus matching the precision requirements of current and forthcoming cosmological surveys. Moreover, when used as input for \texttt{halofit}---a numerical scheme that computes the matter power spectrum on nonlinear scales---our emulator maintains sub-percent errors on average when compared against \class~outputs.

We further extended our framework to incorporate parametric corrections designed to mimic scale-dependent features predicted by modified gravity theories. These extensions preserve the analytical nature of the model and allow us to parametrize characteristic deviations in the MPS due to clustering dark energy or fifth forces, with interpretable parameters such as $\gamma_\MG$, $k_\MG$, and $\sigma_\MG$. When applied to a representative $f(R)$ gravity model, the resulting non-wiggly spectra successfully captured the amplification of power at small-scales. It is worth emphasizing that, in this work, we do not train a general modified-gravity emulator. Instead, we use \texttt{mg\_class} to compute linear spectra for specific models and fit a parametric deformation for the smoothed component, achieving average fractional errors of about 1.5\%. This example serves as an illustration of how the proposed framework can be extended to capture the global modulation of the MPS in alternative gravity models while retaining interpretability.

To assess the observational implications of our model, we computed the two-point correlation function and compared the resulting BAO template with those obtained using conventional de-wiggling techniques. Our formula recovers the BAO peak location with high fidelity across different cosmologies. Moreover, we were able to show that the BAO scale is slightly modified due to deviations from GR as modeled by the Hu-Sawicki model of $f(R)$ theories, in particular, shifting the BAO peak to smaller scales.

A key distinction of our methodology lies in its interpretability: each term in the final expression is rooted in well-understood physical mechanisms, which enables transparent diagnostics and facilitates theoretical extensions—e.g. the corrections added to improve the accuracy at some specific scales. This work demonstrates that symbolic regression—when informed by physical priors—offers a powerful alternative to black-box emulators or semi-analytical expressions that prioritize raw accuracy at the expense of clarity. The resulting formulas are not only compact and fast to evaluate, but also transparent and extendable, making them ideally suited for applications in large-scale structure analyses, parameter inference, and theoretical modeling.

\textbf{Final Remark:} While the final fitting function for the $\Lambda$CDM MPS involves multiple correction terms, its symbolic structure remains compact and physically motivated. In future work, we plan to release a user-friendly implementation of this emulator as a standalone package. This tool will allow users to input cosmological parameters and obtain the corresponding linear MPS with sub-percent accuracy, including all necessary corrections, in a fast and interpretable manner. For the time being, in Appendix~\ref{App: User}, we provide a summary of the formulation as an \textbf{User's Guide}.

\section*{Acknowledgements}

JBOQ would like to express his gratitude to the Facultad de Ciencias F\'isicas y Matem\'aticas of Universidad de Chile for their hospitality during the completion of this work. JBOQ is supported by Patrimonio Aut\'onomo--Fondo Nacional de Financiamiento para la Ciencia, la Tecnolog\'ia y la Innovaci\'on Francisco Jos\'e de Caldas (MINCIENCIAS-COLOMBIA) Grant No. 110685269447 RC-80740-465-2020,
project 69723 and by Vicerrector\'ia de Investigaciones
- Universidad del Valle Grant No. 71357. DS acknowledges support from Fondecyt Regular N.~1251339. SN acknowledges support from the research project PID2021-123012NB-C43 and the Spanish Research Agency (Agencia Estatal de In-
vestigaci\'on) through the Grant IFT Centro de Excelencia Severo Ochoa No CEX2020-001007-S, funded by MCIN/AEI/10.13039/501100011033.

\appendix

\section{User's Guide}
\label{App: User}

This appendix provides a practical guide for users who wish to implement and apply the emulator for the linear matter power spectrum (MPS) developed in this work. The goal is to offer a self-contained reference summarizing the required inputs, the complete set of formulas, the normalization procedure, and several reproducible examples, referencing the numerical coefficients in the main text.

\subsection{Overview of the Emulator}

The emulator presented in this work provides a fast and accurate prediction of the linear MPS, $P(k| \boldsymbol{\theta})$, as a function of five cosmological parameters $\boldsymbol{\theta} = \{h, \omega_b, \omega_m, n_s, A_s\}$, whose ranges are provided in Table~\ref{Tab: Params}. The model decomposes the matter transfer function (MTS) into a smooth (no-wiggle) component, a BAO oscillatory component, and supplements it with a some localized corrections designed to capture detailed features around the equality, the BAO scale, and Silk damping. When tested against spectra retrieved from \class, the emulator achieves a mean absolute percentage error error of $\mape \simeq 0.28\%$ in the range $k = [10^{-5}, 1.5]~h\,\Mpc^{-1}$.

\subsection{De-Wiggled Matter Power Spectrum}

The non-wiggle MPS is given by the closed-form expression:
\begin{equation}
    P_\nw(k) = A_0\, k^{n_s} T^2_{\nw}(k).
\end{equation}
The normalization constant $A_0$ follows from the definition of $\sigma_R$:
\begin{equation}
    \sigma_R^2 = \frac{A_0}{2\pi^2} \int_0^\infty \dd k\, k^{2 + n_s} T^2_\nw(k)\, W^2(kR),
\end{equation}
for $R = 8~\Mpc\,h^{-1}$, using the top-hat filter:
\begin{equation}
    W(x) \equiv \frac{3}{x^3} \left( \sin x - x \cos x \right).
\end{equation}
The smooth MTS is given by:
\begin{equation}
    T_\nw(q) = \left[ 1 + \sum_{i = 1}^{4} a_i q^{b_i}  \right]^{-1/4}, \ q \equiv \frac{k \, h}{\omega_m - \omega_b},
\end{equation}
with coefficients listed in Table~\ref{Tab: Tnw}. This approximation reproduces the \class~MPS with an average fractional error of $\mape(P_\nw) = 0.99\%$.

\subsection{Wiggly Matter Power Spectrum}

The MPS including wiggles reads:
\begin{equation}
    P_\GA(k) \equiv A_0\, k^{n_s}\, T_\GA^2(k),
\end{equation}
where the full MTS is:
\begin{equation}
    T_\GA(k) \equiv T_\nw(k)\, T_\w(k)\, T_{\text{S},1}(k),
\end{equation}
and the local Silk-region correction is:
\begin{equation}
    T_{\S,1}(k) \equiv 1 - A_{\text{S},1} e^{-(k - k_{\text{S},1})^2/\sigma_{\text{S},1}^2},
\end{equation}
with coefficients given in Eq.~\eqref{Eq: Coeff S1}. The wiggly MTS is:
\begin{equation}
    T_\w(k) \equiv \left[1 + f_\text{amp}~e^{-f_\Silk} \sin \left( f_\text{osc} \right) \right].
\end{equation}
The amplitude, Silk damping, and oscillatory phase functions are:
\begin{align}
    f_\text{amp}(k) &\equiv \frac{f_\alpha(\omega_b, \omega_m)}{a_5 + \left\{f_\beta(\omega_b, \omega_m)/(k h s_\GA)\right\}^{b_5}}, \\
    f_\Silk(k) &\equiv a_6(kh/k_\Silk)^{b_6}, \\
    f_\text{osc}(k) &\equiv \frac{a_7 (k h s_\GA + a_8\,\omega_m^{-b_7})}{\left(a_9 + \left\{ f_\text{node}(\omega_m)/(k h s_\GA)\right\}^{b_8} \right)^{b_9}}.
\end{align}
The sound horizon and Silk scales are:
\begin{align}
    s_\GA &\equiv \frac{1}{c_1 \omega_b^{c_2} + c_3\,\omega_m^{c_4} + c_5\omega_b^{c_6}\omega_m^{c_7}}, \\
    k_\Silk &\equiv 0.373\, \omega_b^{0.419} + 0.195\,\omega_m^{1.0957},
\end{align}
with coefficients given in Eq.~\eqref{Eq: Coeff sGA}. The remaining auxiliary functions are
\begin{align}
    f_\alpha (\omega_b, \omega_m) &\equiv a_{10} - a_{11} \omega_b^{b_{10}} + a_{12} \omega_m^{b_{11}}, \\
    f_\beta (\omega_b, \omega_m) &\equiv b_{12} - a_{13} \omega_b^{b_{13}} + a_{14} \omega_m^{b_{14}}, \\
    f_\text{node} (\omega_m) &\equiv a_{15} \omega_m^{b_{15}}.
\end{align}
All the coefficients for $T_\w$ are listed in Table~\ref{Tab: Hyperparameters}. The average error of the wiggly spectrum is $\mape(P_\GA)=0.42\%$.

\subsection{Corrections around the Silk Scale}

The BAO–Silk transition is further refined with three localized Gaussian corrections:
\begin{equation}
    P_{\S,i}(k) \equiv 1 + A_{\S,i}\,e^{-(k - k_{\S,i})^2/\sigma_{S,i}^2},
\end{equation}
yielding:
\begin{equation}
    P_\GA^\text{corr}(k) \equiv P_\GA(k) \times P_{\S,2}(k) \times P_{\S,3}(k) \times P_{\S, 4}(k),
\end{equation}
with coefficients given below Eq.~\eqref{Eq: PGA Silk Corrections}. The corrected spectrum reproduces \class~up to $\mape = 0.39\%$. $\mape(P_\GA^{(\text{corr})}) = 0.39\%$.

\subsection{Corrections around the Equality Scale}

Near the equality peak we introduce a skew–normal correction:
\begin{align}
    &P_\max(k) \equiv 1 \nonumber \\
    &+ \left\{A_\max(k - k_\max) + B_\max\right\} 
    e^{-\frac{1}{2}(k - k_\max)^2/\sigma_\max^2} \nonumber \\
    &\times \left[1 + \text{Erf}\left(\lambda_\max \frac{k - k_\max}{\sqrt{2}\sigma_\max}\right) \right]. 
\end{align}
The full MPS is thus:
\begin{equation}
    P_\text{full}(k) \equiv P_\GA^{(\text{corr})}(k) \times P_\max(k)
\end{equation}
The parameters follow:
\begin{align}
    B_\max &= R_\max - 1, \\
    A_\max = &-\left( \frac{P'_{\GA, \max}}{P_{\GA,\max}} \right)R_\max - \sqrt{\frac{2}{\pi}} \left( \frac{\lambda_\max}{\sigma_\max} \right) B_\max, 
\end{align}
with:
\begin{align}
    R_\max &= 0.6461 - 0.0097 h^2 + 0.0307 n_s \nonumber \\
          &\quad + 7.1728 \omega_b + 0.0239\, \omega_m^{-1}.
\end{align}
\begin{align}
    \frac{P_\GA(k_\max)}{A_0} &= 0.047\, \frac{\omega_m^{1.06224}\, e^{-8.50419\omega_b}}{h^{0.89981}\, n_s^{3.91548}}, \\
    \frac{P'_\GA(k_\max)}{A_0} &= -0.40649 + 0.00518\, h - 0.00139\, h^2 \nonumber \\
    &\quad + 0.26388\, n_s - 0.10340\, n_s^2 \nonumber \\
    &\quad - 2.22729\, \omega_b - 18.4715\, \omega_b^2 \nonumber \\
    &\quad + 3.0768\, \omega_m - 6.70755\, \omega_m^2,
\end{align}
\begin{align}
   \log_{10} \left[ P''_\class(k_\max) \right] &= 
   10.9041 - \cos h - \frac{1.1084}{h + 3\omega_b} + \omega_b \nonumber \\
   &+ \log\left( \frac{(A_s/10^9)^{0.4338}}{n_s} \right),
\end{align}
and:
\begin{equation}
    k_\max = \frac{0.07066 \, \omega_m^{0.8824} \, n_s^{0.939}}{h^{1.006} (1 + 1.2025 \, \omega_b)^{3.3395}}~[h\,\Mpc^{-1}].
\end{equation}
We fix $\lambda_\max=0.4$ empirically, and determine $\sigma_\max$ from the condition:
\begin{equation}
    \left. \frac{\dd^2}{\dd k^2} P_\class\right|_{k_\max} 
    = \left. \frac{\dd^2}{\dd k^2} \left( P_\GA \, P_\max \right) \right|_{k_\max}.
\end{equation}
The final full spectrum achieves an average fractional error of $\mape(P_\text{full}) = 0.28\%$. \\

\subsection*{Code Availability and Algorithm Settings}

A public implementation in \texttt{Mathematica} and \texttt{Python} is available at:
\begin{center}
\href{https://github.com/BayronO/Pk-Emulator-from-GAs}{\texttt{https://github.com/BayronO/Pk-Emulator-from-GAs}}
\end{center}
The repository includes all scripts and data used in this work, together with a compact tutorial illustrating how to evaluate each component of the emulator. 

However, to facilitate the reproduction of our training pipeline, we summarize the specific configurations of the solvers used.

\textbf{Genetic Algorithm:} We employed a custom GA to perform symbolic regression on the transfer functions. Crucially, the functional form of the search was restricted to polynomial structures generated via the grammar $G = \{+, -, \times, \div\}$ acting on the variables $\{k, h, \omega_b, \omega_m\}$. These polynomials serve as arguments to a fixed ansatz containing exponential and sinusoidal modulation [as detailed in Eq.~\eqref{Eq: TGA_nw}]. The algorithm optimizes the coefficients and exponents within this fixed structure. To prevent bloat (overfitting), parsimony pressure is enforced by using a fixed-length matrix encoding which sets a hard upper limit on the expression complexity. The hyperparameters governing the evolution are listed in Table~\ref{Tab: GA Settings}.

\begin{table}[h]
    \centering
    \caption{Hyperparameters for the GA. The chromosome structure refers to the fixed-size matrix encoding (genes $\times$ nucleotides) used to enforce parsimony.}
    \label{Tab: GA Settings}
    \setlength{\tabcolsep}{4pt}
    \renewcommand{\arraystretch}{1.2}
    \begin{tabular}{l r}
    \hline
    \hline
    \textbf{Parameter} & \textbf{Value} \\
    \hline
    \multicolumn{2}{c}{\textbf{Global Settings}} \\
    Population Size & 100 \\
    Mutation Rate & 0.35 \\
    Selection Scheme & Tournament (size 4) \\
    Selection Rate & 0.3 (replacement) \\
    \hline
    \multicolumn{2}{c}{\textbf{Smooth Function ($T_\nw$)}} \\
    Crossover Rate & 0.75 \\
    Structure & $6 \times 6$ matrix \\
    \hline
    \multicolumn{2}{c}{\textbf{Wiggly Function ($T_\w$)}} \\
    Crossover Rate & 0.70 \\
    Structure & $10 \times 10$ matrix \\
    \hline
    \hline
    \end{tabular}
\end{table}

\textbf{Boltzmann Solver:} The spectra for testing were generated using the \class~code (version 3.2.3) interfaced with the modified gravity patches \texttt{mg\_class} (version 2.9.4, configured for \texttt{mg\_ansatz} $=$ \texttt{plk\_late} and \texttt{FR}) and \texttt{hi\_class} (version 3.2.3). These spectra were retrieved for 200 logarithmically spaced values covering the range $k \in [10^{-4}, 1.5]~h\,\Mpc^{-1}$. We computed the non-linear MPS at $z=0$ using the \texttt{halofit} module with default precision settings. In this case, the integration was extended to $k_{\text{max}} = 10~h\,\Mpc^{-1}$. To efficiently span the parameter space, the test set was sampled using a Latin Hypercube generator (via \texttt{scipy.stats.qmc}) with the random seed fixed to 3.

\textbf{PySR Configuration:} For the regression of $P''(k_\max)$, we utilized the \texttt{PySR} package (interfaced with the Julia backend). The symbolic search was conducted over 5000 iterations using a grammar consisting of the binary operators $\{+, -, \times, /, \text{pow}\}$ and the unary operator $\{\exp\}$. The optimization objective was to minimize the Mean Squared Error (MSE) on the $\log_{10}$-transformed target variable. To enforce parsimony and prevent overfitting, we imposed a hard complexity limit of 50 nodes per expression (\texttt{maxsize=50}). The final model was chosen using the \texttt{model\_selection="best"} strategy, which selects the expression that maximizes accuracy while penalizing excessive complexity.

\section{Improved Formulas from GAs}
\label{App: GA formulas}

In this section, we present improved analytic formulas for two physically relevant scales in cosmology: the scale at the $P(k)$ maximum $k_\max$ and the photon diffusion (Silk) scale $k_\Silk$. These expressions are derived from the same set of spectra used to train our GA-based model and are shown to outperform the widely used EH approximations in accuracy.

\subsection{Scale at the MPS Maximum $k_\max$}

Physically, the location of the turnover in the linear MPS—near to the scale of equality $k_\eq$ but not exactly equal—is influenced by several cosmological parameters. In particular, we find it depends non-negligibly on $h$, $n_s$, and weakly on $\omega_b$, while $A_s$ affects only the amplitude of this maximum and not its scale.

By computing the actual location of the peak of the MPS used in training our GA model, we empirically determine the following fit:
\begin{equation}
    k_\max = \frac{0.07066 \, \omega_m^{0.8824} \, n_s^{0.939}}{h^{1.006} (1 + 1.2025 \, \omega_b)^{3.3395}}~[h\,\Mpc^{-1}].
\end{equation}
When tested against the $k_\max$-values obtained from the 200 spectra in the test set, this formula yields a remarkable accuracy with an average fractional error of $\mape(k_\max) = 0.038\%$.

\subsection{Photon Diffusion (Silk) Scale \(k_\Silk\)}

The Silk damping scale corresponds to the suppression of power on small scales due to photon diffusion prior to recombination. The EH fitting formula for this scale is~\cite{Eisenstein:1997ik}:
\begin{equation}
\label{Eq: kSilk EH}
    k_\Silk^{(\EH)} = 1.6 \, \omega_b^{0.52} \, \omega_m^{0.73} \left[1 + \left( 10.4 \, \omega_m \right)^{-0.95} \right],
\end{equation}
in units of $[1/\Mpc]$. 

Physically, $k_\Silk$ corresponds to the inverse of the comoving diffusion length of photons at the time of recombination. We compute this quantity numerically using the thermodynamics module in \class, which returns the photon damping wavenumber \texttt{kd} in units of $[1/\Mpc]$. Since $k_\Silk$ is insensitive to $n_s$, $h$, or $A_s$, we vary only $\omega_b$ and $\omega_m$ to obtain a reliable fit.

The following expression provides an excellent approximation:
\begin{equation}
    k_\Silk = 0.373\, \omega_b^{0.419} + 0.195\,\omega_m^{1.0957}~[h/\Mpc],
\end{equation}
with a mean fractional error of only $\mape(k_\Silk) = 0.03\%$, compared to $\mape(k_\Silk^{(\EH)}) = 36\%$ for the EH approximation.

\section{Full EH Fitting Formula}
\label{App: Full EH}

The transfer function given by Eisenstein and Hu \cite{Eisenstein:1997ik} has the following form:
\begin{equation}
T(k) = \frac{\Omega_b}{\Omega_0} T_b(k) + \frac{\Omega_c}{\Omega_0} T_c(k),
\end{equation}
where $\Omega_0 = \Omega_b + \Omega_c$. The terms involved in this formula are the following:
\begin{equation}
\label{Eq: Tb EH}
    T_b = \Bigg[\frac{\tilde{T}_0(k; 1,1)}{1 + \left( \frac{ks}{5.2} \right)^2} + \frac{\alpha_b}{1 + \left(\frac{\beta_b}{ks} \right)^3} e^{-\left( \frac{k}{k_{\mathrm{Silk}}} \right)^{1.4}}\Bigg] j_0(k \tilde{s}),
\end{equation}

\begin{align}
    T_c &= f \tilde{T}_0(k, 1, \beta_c) + (1 - f) \tilde{T}_0(k, \alpha_c, \beta_c), \\
    \tilde{T}_0&(k, \alpha_c, \beta_c) = \frac{ \ln(e + 1.8 \beta_c q)}{\ln(e + 1.8 \beta_c q) + C q^2}, \\
    \nonumber
    \end{align}
    \begin{align}
    f &= \frac{1}{1+(ks/5.4)^4}, \\
    C &= \frac{14.2}{\alpha_c} + \frac{386}{1 + 69.9 q^{1.08}}, \\
    q &= \frac{k}{13.41 k_{\mathrm{eq}}}, \\
    R &\equiv 3\rho_b/4\rho_\gamma = 31.5 \omega_b \Theta_{2.7}^{-4} (z/10^3)^{-1}, \\
    \nonumber
\end{align}

\begin{align}
    k_{\mathrm{Silk}} &= 1.6 \omega_b^{0.52} \omega_0^{0.73} \left[1 + \left( 10.4 \, \omega_0 \right)^{-0.95} \right] \,  \mathrm{Mpc}^{-1}, \\
    k_{\mathrm{eq}} &= 7.46 \times 10^{-2} \omega_0 \Theta_{2.7}^{-2} \, \mathrm{Mpc}^{-1}, \\
    \nonumber
\end{align}

\begin{align}
    \alpha_b &= 2.07 k_{\mathrm{eq}} s (1 + R_d)^{-3/4} G \left( \frac{1 + z_{\mathrm{eq}}}{1 + z_d} \right), \\
    \beta_b &= 0.5 + \frac{\Omega_b}{\Omega_0} + \left( 3 - 2 \frac{\Omega_b}{\Omega_0} \right) \sqrt{\left( 17.2 \, \omega_0 \right)^2 + 1}, \\
    \nonumber
\end{align}

\begin{align}
    \beta_{\mathrm{node}} &= 8.41 \omega_0^{0.435}, \\
    s &= \frac{2}{3 k_{\mathrm{eq}}} \sqrt{\frac{6}{R_{\mathrm{eq}}}} \ln \frac{\sqrt{1 + R_d} + \sqrt{ R_d +   R_{\mathrm{eq}}}}{1 + \sqrt{R_{\mathrm{eq}}}}, \\
    \tilde{s} &= s \left[1 + \left( \frac{\beta_{\mathrm{node}}}{k s} \right)^3 \right]^{-1/3}, \\
    \nonumber
\end{align}

\begin{align}
    G(y) &= - 6 y \sqrt{1 + y} \nonumber \\
    &+ y (2 + 3y) \ln \left(\frac{
    \sqrt{1 + y} + 1}{\sqrt{1 + y} -1} \right), \\
    y &\equiv \frac{1 + z_\text{eq}}{1 + z}, \\
    \nonumber
\end{align}

\begin{align}
    \alpha_c &= a_1^{-\Omega_b/\Omega_0} a_2^{-(\Omega_b/\Omega_0)^3}, \\
    a_1 &= (46.9 \omega_0)^{0.670} [ 1 + (32.1 \omega_0)^{-0.532}], \\
    a_2 &= (12.0 \omega_0)^{0.424} [1 + (45.0 \omega_0 )^{-0.582}], \\
    \nonumber
\end{align}

\begin{align}
    \beta_c^{-1} &= 1 + b_1 [(\Omega_c/\Omega_0)^{b_2} - 1], \\
    b_1 &= 0.944 [ 1 + (458 \omega_0)^{-0.708} ]^{-1}, \\
    b_2 &= (0.395 \omega_0)^{-0.0266}, \\
    \nonumber
\end{align}

\begin{align}
    z_{\mathrm{eq}} &= 2.50 \times 10^4 \omega_0 \Theta_{2.7}^{-4}, \\
    z_d &= 1291 \frac{\omega_0^{0.251}}{1 + 0.659 \omega_0^{0.828}} \left[ 1 + b_{1, z} \omega_b^{b_{2, z}} \right], \\
    b_{1, z} &= 0.313 \omega_0^{-0.419} \left[ 1 + 0.607 
    \omega_0^{0.674} \right], \\
    b_{2, z} &= 0.238 \omega_0^{0.223},
\end{align}
where it has been defined $\omega_0 = (\Omega_c + \Omega_b) h^2$, $T_\text{CMB} \equiv 2.7 \Theta_{2.7} \, \text{K}$, $R_d \equiv R(z_d)$ $R_\text{eq} \equiv R(z_\text{eq})$.

\bibliographystyle{utcaps.bst} 
\bibliography{Utils/Bibli.bib}

\end{document}